\documentclass[11pt]{article}
\usepackage{jheppub}
\usepackage{times}
\usepackage{graphicx}
\usepackage{amssymb}
\usepackage{amsmath}
\newcommand{\be}{\begin{equation}}  \newcommand{\ee}{\end{equation}\noindent}
\newcommand{\bea}{\begin{eqnarray}} \newcommand{\eea}{\end{eqnarray}}
\newcommand{\nn}{\nonumber}
\newcommand{\bc}{\begin{center}}  \newcommand{\ec}{\end{center}}
          
\newcommand{\calD}{{\cal D}}      
\newcommand{\maprightb}[1]{\smash{\mathop{
\hbox to 1cm{\rightarrowfill}}\limits_{#1}}}
\def\del{\partial}
   
\newcommand{\bra}{\langle}     \newcommand{\ket}{\rangle}

\newcommand{\Tpc}{T_{\rm pc}}    \newcommand{\betapc}{\beta_{\rm pc}}
\newcommand{\Csw}{C_{\rm SW}}    
\newcommand{\Nred}{ {N_{\rm red}} }

 \newcommand{\Rbar}{\bra R\ket_0}   
\newcommand{\ketSP}{\rangle_0}
\newcommand{\RRW}{R} 

\begin{document}
\title{
EoS of finite density QCD with Wilson fermions by Multi-Parameter 
Reweighting and Taylor expansion
}
\author{Keitaro Nagata and Atsushi Nakamura}
\affiliation{Research Institute for Information Science and Education, 
Hiroshima University, \\
Higashi-Hiroshima 739-8527 Japan}
\emailAdd{kngt@hiroshima-u.ac.jp}
\emailAdd{nakamura@riise.hiroshima-u.ac.jp}
\date{\today}
\abstract{
The equation of state (EoS), quark number density and susceptibility
at nonzero quark chemical potential $\mu$ are studied in lattice QCD simulations  
with a clover-improved Wilson fermion of $2$-flavors and RG-improved gauge action.
To access nonzero $\mu$, we employ two methods : a multi-parameter reweighting (MPR) in 
$\mu$ and $\beta$ and Taylor expansion in $\mu/T$. 
The use of a reduction formula for the Wilson fermion
determinant enables to study the reweighting factor in
MPR explicitly and higher-order coefficients in
Taylor expansion free from errors of noise method, 
although calculations are limited to small lattice size.
As a consequence, we can study the reliability of the thermodynamical quantities 
through the consistency of the two methods, each of which has different origin of the application limit.

The thermodynamical quantities are obtained from simulations on a $8^3\times 4$ 
lattice with an intermediate quark mass($m_{\rm PS}/m_{\rm V}=0.8)$. 
The MPR and Taylor expansion are consistent for the EoS and number density up 
to $\mu/T\sim 0.8$ and for the number susceptibility up to $\mu/T \sim 0.6$. 
This implies within a given statistics that the overlap problem for the MPR and 
truncation error for the Taylor expansion method are negligible in these regions.

In order to make MPR methods work, the fluctuation of the reweighting factor should be small.
We derive the equation of the reweighting line where the fluctuation is small,  
and show that the equation of the reweighting line is consistent with the 
fluctuation minimum condition. 
}
\maketitle

\section{Introduction}
Thermodynamical properties of strongly interacting matter have been of prime interest
in hadron physics. Such an understanding is inevitable to complete the
understanding of states of matter such as normal nuclear matter, 
quark-gluon plasma and dense matters, which are related to the 
study of evolution of universe, heavy ion collisions, and dense matter inside compact stars. 

Lattice QCD is a powerful method to study the non-perturbative nature of QCD. 
However, the introduction of quark chemical potential $\mu$ causes the sign problem for 
lattice QCD simulations, and standard Monte Carlo(MC) techniques are not applicable for 
$\mu\neq 0$~\cite{deForcrand:2010ys}.  
Several methods have been developed to deal with nonzero-$\mu$ systems in 
lattice QCD simulations~\cite{Muroya:2003qs,Schmidt:2006us,deForcrand:2010ys}. 

A reweighting is a general technique for MC simulations to 
reduce numerical costs~\cite{Ferrenberg:1988yz}. 
Let us consider a space spanned by parameters of a system. 
An idea of the reweighting is to perform importance sampling at a point on the 
parameter space (simulation point), and to calculate observables for other points 
(target point) by using the samples obtained at the simulation point. 
The reweighting provides a reweighting factor to compensate the difference of weights between 
the two points. 
This method was applied for chemical potential in the Glasgow 
method~\cite{Barbour:1991vs,Barbour:1997ej}. 
Later Fodor and Katz proposed a method to improve the reweighting method 
by adopting multiple parameters as shifted parameters~\cite{Fodor:2001au}, 
which is referred to as the multi-parameter reweighting (MPR) method. 
The location of the critical end point  and the equation of state was investigated,  
by using the MPR method with staggered fermions with four-flavor~\cite{Fodor:2001au} and 
2+1 flavor~\cite{Fodor:2001pe,Fodor:2002km,Fodor:2004nz}.  
See also Ref.~\cite{Fodor:2009ax}. 

Although MPR provides a way to investigate QCD at $\mu\neq 0$ by 
circumventing the breakdown of MC methods, it may encounter problems
caused by the fluctuation of the reweighting factor. 
The fluctuation of the phase of the reweighting factor causes the sign oscillation appearing at the step of the ensemble 
average of observables. Large phase-fluctuation makes MPR unreliable~\cite{Ejiri:2004yw}. 
On the other hand, large fluctuation of the absolute value causes the decrease
of the number of effective samples, which implies less overlap between important 
configurations at the simulation point and and those at the target point. 

Another approach to study QCD at $\mu \neq 0$ is to make use of the Taylor expansion 
at $\mu=0$, which has been studied in e.g. Refs.~\cite{Allton:2002zi,Allton:2003vx,Allton:2005gk,Gavai:2004sd,Ejiri:2009hq}.
The use of the Taylor expansion methods needs a careful investigation 
on the truncation error of the Taylor series, especially for near and below the pseudo critical 
temperature $\Tpc$. 

The two approaches suffer from different systematic errors: 
the overlap and sign problems for MPR, and the truncation errors for the Taylor expansion. 
Therefore, it is valuable to study their consistency, which provides 
a complementary way to confirm the reliability of calculations. 

In the present work, we calculate thermodynamical quantities by using MPR 
and Taylor expansion with a careful attention on their consistency.
Although the consistency is empirically known, it is important to 
show the consistency explicitly in a way free from statistical 
errors such as noise or truncation errors of Taylor expansion. 
  
We also investigate the validity of MPR. The validity of the MPR method were 
investigated in detail in Refs.~\cite{Csikor:2004ik,Ejiri:2004yw} by using staggered fermions. 
The fermion determinant controls the phase fluctuation of the reweighting factor. 
Hence, the numerical difficulty of MPR is caused in part by the fluctuation of the fermion determinant. 
In addition, reweighting lines depend on the parameters of the actions. 
Hence, it is important to investigate MPR by different fermion actions.

For the purpose, we evaluate the fermion determinant exactly with the use 
of a reduction formula for Wilson fermions~\cite{Borici:2004bq,Nagata:2010xi,Alexandru:2010yb}. 
As we will see later, the formula makes it feasible to evaluate the determinant 
without any approximation. 
In addition, the formula describes the quark determinant as an analytic function of $\mu$. 
This feature enables to evaluate the determinant for an arbitrary value of $\mu$, and 
makes it easy to evaluate higher-order Taylor coefficients. 
However, note that the determinant evaluation needs large numerical cost even though 
the reduction formula is used, which imposes the limitation on the applicable lattice 
size.

This paper is organized as follows. We explain the framework in the next section. 
The MPR method is introduced in \ref{subsec2a}, the overlap problem and 
the reweighting line to suppress the overlap problem is discussed in \ref{subsec2b}. 
The reduction formula is presented in \ref{subsec2c}. 
Numerical results are shown in section~\ref{sec3}. 
Simulation setup is given in ~\ref{subsec3a}. Properties of the 
fermion determinant and reweighting factor is investigated in ~\ref{subsec3b}
and \ref{subsec3c}. Then, the consistency of MPR and Taylor expansion 
for EoS et. al. is discussed in ~\ref{subsec3d}. 
We also make a comparison with imaginary chemical potential approach
in ~\ref{subsec3e}. Finite size effect on MPR is mentioned in 
~\ref{subsec3f}. 
The final section is devoted to a summary.

\section{Framework}
\label{sec2}
\subsection{Action and thermodynamical quantities}
The grand partition function of $N_f$-flavor QCD at a temperature $T$ and quark chemical potential $\mu$ is given by 
\begin{align}
Z_{GC}(\mu, T) = \int \calD U \;[\det \Delta(\mu)]^{N_f} e^{-\beta S_G}.
\label{Sep122011eq1}
\end{align}
Here $S_G$ is the RG-improved gauge action divided by $\beta$. 
$N_f$ is the number of the flavors, where we consider $N_f=2$ in simulations. 
This definition of $S_G$ is convenient in the MPR method. 
We employ the clover-improved Wilson fermion 
\begin{align}
\Delta(\mu) =  \delta_{x, x^\prime} &-\kappa \sum_{i=1}^{3} \left[
(1-\gamma_i) U_i(x) \delta_{x^\prime, x+\hat{i}} 
+ (1+\gamma_i) U_i^\dagger(x^\prime) \delta_{x^\prime, x-\hat{i}}\right] \nonumber \\
 &-\kappa \left[ e^{+\mu a} (1-\gamma_4) U_4(x) \delta_{x^\prime, x+\hat{4}}
+e^{-\mu a} (1+\gamma_4) U^\dagger_4(x^\prime) \delta_{x^\prime, x-\hat{4}}\right] \nonumber \\
&- \kappa  C_{SW} \delta_{x, x^\prime}  \sum_{\mu \le \nu} \sigma_{\mu\nu} 
F_{\mu\nu},
\label{Jul202011eq1}
\end{align}
where  $\kappa$ and $C_{SW}$ are the hopping parameter and clover coefficient.
In a homogeneous system, the EoS at $T$ and $\mu$ is defined by 
$p=(T/V_s)\ln Z_{GC}$, which is
\begin{align}
\frac{p(\mu, T)}{T^4} = \left(\frac{N_t}{N_s}\right)^3 \ln Z_{GC} (\mu, T)
\label{Sep122011eq2}
\end{align}
in the lattice with spatial extent $N_s (=N_x=N_y = N_z)$ and temporal extent $N_t$.
On this lattice $T=(a N_t)^{-1}$ and $V_s=(a N_s)^3$ with a lattice spacing $a$. 
In simulations, we consider the deviation of the pressure from $\mu=0$
\begin{subequations}
\begin{align}
\frac{\delta p(\mu, T)}{T^4} = \frac{p(\mu,T)}{T^4} - \frac{p(0,T)}{T^4} _.
\end{align}
The quark number density and quark number susceptibility are given by 
\begin{align}
\frac{n}{T^3} & = \frac{ \del}{\del (\mu/T)} \frac{\delta p}{T^4} \nn \\
              & = \left(\frac{N_t}{N_s}\right)^3 \left\langle  \frac{  (T \del/\del\mu) [\det \Delta(\mu)]^{N_f}}{[\det \Delta(\mu)]^{N_f}} \right\rangle_, \\
\frac{\chi}{T^2} &= \frac{ \del^2}{\del (\mu/T)^2} \frac{\delta p}{T^4} \nn \\ 
  & =\left(\frac{N_t}{N_s}\right)^3 \left[ 
  \left\langle  \frac{  (T \del/\del\mu)^2 [\det \Delta(\mu)]^{N_f}}{[\det \Delta(\mu)]^{N_f}} \right\rangle -
\left\langle  \frac{  (T \del/\del\mu) [\det \Delta(\mu)]^{N_f}}{[\det \Delta(\mu)]^{N_f}} \right\rangle^2 \right]_. 
\end{align}
\label{Nov092011eq1}
\end{subequations}

\subsection{Multi-parameter reweighting}
\label{subsec2a}
\def\rwfac{ \left(\frac{\det \Delta(\mu) }{ \det \Delta(0)}\right)^{N_f} e^{-(\beta-\beta_0) S_G} }
To calculate Eqs.~(\ref{Nov092011eq1}) for $\mu\neq 0$, 
we employ the MPR method regarding $\mu$ and $\beta$~\cite{Fodor:2001au,Csikor:2004ik}.

The Boltzmann weight 
\begin{align}
w(\mu, \beta) = [\det \Delta(\mu)]^{N_f} e^{- \beta S_G},
\end{align}
provides a probability in importance sampling.  
However, it is unfeasible to update gauge configurations with $w(\mu, \beta)$ for 
$\mu\neq 0$, because it is in general complex. 
A basic idea of MPR is to decompose $w(\mu, \beta)$ into two parts as
\begin{align}
w(\mu, \beta) = \RRW(\mu, \beta)_{(0,\beta_0)} \; w(0, \beta_0), 
\label{Oct182011eq1}
\end{align}
and to perform importance sampling at $(0, \beta_0)$ with 
$w(0, \beta_0)$ as the probability. 
The remaining factor $\RRW(\mu,\beta)_{(0,\beta_0)} \equiv w(\mu, \beta)/w(0, \beta_0) $  
is instead taken into account at the step of the calculation of observables. 
$\RRW$ is often called the reweighting factor and reads 
\begin{align}
\RRW(\mu,\beta)_{(0,\beta_0)}  = \left( \frac{\det \Delta(\mu)}{\det \Delta(0)}\right)^{N_f} e^{ - (\beta -\beta_0) S_G}.
\label{Dec082011eq1}
\end{align}
Note that $\RRW(\mu,\beta)_{(0, \beta_0)}$ is given in terms of configurations 
obtained at $(0, \beta_0)$. The grand-partition function is rewritten as
\begin{align}
Z_{GC}(\mu, T) & = \int \calD U  \rwfac \;[\det \Delta(0)]^{N_f} e^{- \beta_0 S_G}, \nn \\
&= \int \calD U \RRW(\mu, \beta)_{(0,\beta_0)} \; w(0, \beta_0). 
\end{align}
The expectation value of an observable $O$ is given by 
\begin{align}
\bra O \ket & =\frac
{\int \calD U \; O \; \RRW(\mu, \beta)_{(0,\beta_0)} w(0, \beta_0)}
{\int \calD U \; \RRW(\mu, \beta)_{(0,\beta_0)} w(0, \beta_0)} \nn \\
&= \frac{\bra O \;\RRW(\mu, \beta)_{(0,\beta_0)} \ket_0 }{\bra  \RRW(\mu, \beta)_{(0,\beta_0)} \ket_0}.
\label{Oct052011eq1}
\end{align}
Here $\bra \cdot \ket_0$ denotes an average taken over an ensemble generated with 
the importance sampling with the weight $w(0, \beta_0)$. 

In the calculation of the reweighting, it is possible to combine several
ensembles obtained from different parameter sets, for instance
multi-ensemble reweighting~\cite{Ferrenberg:1989ui,Kratochvila:2005mk,deForcrand:2006ec} or\
 multi-histogram method~\cite{Nakagawa:2011eu}.
Although those elaborated techniques are favorable to achieve better overlap,
the reweighting with single ensemble is visible to understand the
consistency between the Taylor expansion and reweighting.
Thus, we use single ensemble reweighting for one target point.

The pressure is given by 
\begin{align}
\frac{\delta p}{T^4} = \left(\frac{N_t}{N_s}\right)^3 \ln
\frac{\bra \RRW(\mu, \beta)_{(0,\beta_0)} \ket_0 }{\bra  \RRW(0, \beta)_{(0,\beta_0)} \ket_0}.
\end{align}
The quark number density and susceptibility are obtained from Eqs.~(\ref{Nov092011eq1}) and 
(\ref{Oct052011eq1}). 

\subsection{Overlap problem and reweighting line}
\label{subsec2b}

The expectation value $\bra O\ket$ given in Eq.~(\ref{Oct052011eq1}) would be 
independent of the location of the simulation point $(0,\beta_0)$ in the 
parameter space, if a sufficiently large number of measurements is considered. 
In practice, $\bra O\ket$ depends on $(0, \beta_0)$ in simulations with a 
finite number of samples. 
The problem arises from the fluctuation of the reweighting 
factor~\cite{Csikor:2004ik,Ejiri:2004yw}. 
It was found~\cite{Fodor:2001pe} that a better overlap can be obtained by using 
multiple parameters as reweighting parameters and by changing them appropriately.

Let $X$ the fluctuation of the reweighting factor, 
\begin{align}
X(\mu, \beta) = \langle (\RRW - \bra \RRW\ketSP)^2\ketSP. 
\label{Dec082011eq2}
\end{align}
Here we suppress the arguments of $\RRW(\mu, \beta)_{(0,\beta_0)}$ and 
describe it as $\RRW$. 
The condition for the parameter change is to keep the fluctuation $X$ small. 

Under the change of the parameters $(\mu, \beta)\to (\mu+\Delta \mu, \beta + \Delta \beta)$ with a fixed $\beta_0$, the change of $X$ is given as 
$\delta X = \bra \RRW \delta \RRW  \ketSP   - \Rbar \bra \delta \RRW  \ketSP$.
The fluctuation minimum condition $\delta X =0$ gives 
\begin{align}
\frac{ \bra \RRW \delta \RRW  \ketSP}{\Rbar}   = \bra \delta \RRW  \ketSP.
\label{Nov302011eq3}
\end{align}
By definition, the left hand side is given by 
\begin{subequations}
\begin{align}
\frac{ \bra \RRW \delta \RRW  \ketSP}{\Rbar} = \bra \delta \RRW \ket 
= \frac{1}{Z}\int {\cal D} U \delta \RRW \;w(\mu, \beta), 
\end{align}%
and the right hand side is given by 
\begin{align}
\bra \delta \RRW  \ketSP = \frac{1}{Z}\int {\cal D}U \delta \RRW \; w(0, \beta_0).
\end{align}
\end{subequations}
Equation (\ref{Nov302011eq3}) is satisfied  if the two weights are equal, $w(\mu, \beta)=w(0, \beta_0)$.
This is realized if the target point and simulation point are coincide or if 
the number of configurations are sufficiently large.
\begin{figure}
\begin{center}
\includegraphics[width=7cm]{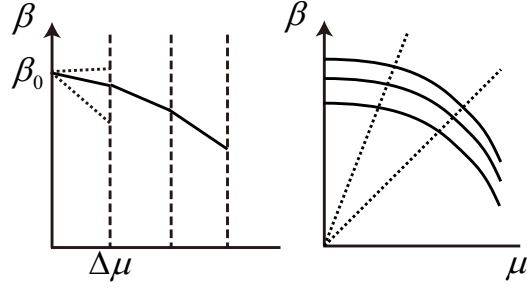}
\caption{The determination of reweighting line and calculation 
of observables such as EoS. 
Left: First, the simulation point $(0, \beta_0)$ is fixed. 
The value of $\beta$ minimizing $X$ is determined for each $\mu$.
Right : Results from several reweighting 
lines are collected to obtain thermodynamical quantities for a given $\mu/T$. 
}\label{Nov132011fig1}
\end{center}
\end{figure}
Instead of the global minimum, we choose the value of $\beta$ that minimizes $X$ for 
each value of $\mu$. The procedure is illustrated in Fig.~\ref{Nov132011fig1}. 
It was pointed out in Ref.~\cite{Ejiri:2004yw} that the phase fluctuation of $R$
can not be canceled by MPR procedure, because the gauge part of $R$ is real. 
In this work, we limit our analysis to regions where the phase fluctuation is small.

The determination of the reweighting line requires the determinant evaluation for 
many parameter sets.  
In the present work, the use of the reduction formula makes this procedure
easier. However, it would be useful to derive an easier way to find the reweighting line.
The deviation of the reweighting factor $\delta \RRW$ for small $\Delta \mu$ and $\Delta \beta$ is given by  
\begin{align}
\delta \RRW(\mu, \beta) = \frac{\del \RRW}{\del (\mu/T)}  \frac{\Delta \mu}{T}
+ \frac{\del \RRW }{\del \beta } \Delta \beta.
\label{Nov302011eq2}
\end{align}
Substitution of Eq.~(\ref{Nov302011eq2}) into Eq.~(\ref{Nov302011eq3}) gives
\begin{align}
\left( \left\langle \frac{\del \RRW}{\del (\mu/T)}\right\rangle - 
\left\langle \frac{\del \RRW}{\del (\mu/T)}\right\rangle_0\right)\frac{\Delta \mu}{T} = - \left( \left\langle\frac{\del \RRW }{\del \beta } \right \rangle - \left\langle\frac{\del \RRW}{\del \beta } \right \rangle_0  \right) \Delta \beta _.
 \label{Nov302011eq0}
\end{align}
This gives the reweighting line. 
Note that $\bra \cdot \ket$ is replaced with $\bra \cdot \ket_0$ according to 
Eq.~(\ref{Oct052011eq1}). 
It can be simplified further 
\begin{subequations}
\begin{align}
\Delta \beta = 
 \frac{ \bra \RRW^2 a\ketSP  - \bra \RRW \ketSP\bra \RRW a\ketSP }{
 \bra \RRW^2 b\ketSP  - \bra \RRW \ketSP\bra \RRW b \ketSP }  \frac{\Delta \mu }{T}_,  
 \label{Nov302011eq1}
\end{align}%
where  
\begin{align}
a & =  \frac{ T \frac{\del }{\del \mu} [\det \Delta(\mu)]^{N_f}}{[\det \Delta(\mu)]^{N_f}}, \\
b & =  S_G.
\end{align}
\end{subequations}
Here we neglect a quark contribution to $b$: $\del \Csw /\del \beta$. 

To find the reweighting line, one can use the equation of the reweighting line 
Eq.~(\ref{Nov302011eq0}) or (\ref{Nov302011eq1}) instead of 
calculating the fluctuation $X$ for many parameter sets. 

It was suggested in ~\cite{Ejiri:2004yw} that the equation of the 
reweighting line may correspond to the Clausius-Clapeyron equation in $(p,T)$ plane. 
Equation (\ref{Nov302011eq1}) has a similar correspondence. Especially, 
it is reduced to $\Delta \beta = ( \bra n \ket - \bra n \ketSP)/(\bra S_G \ket 
- \bra S_G\ketSP)  (\Delta \mu/T)$ in the vicinity of the simulation point.

\subsection{Reduction formula for the Wilson fermion determinant}
\label{subsec2c}

To consider the fluctuation minimum condition, we evaluate 
the quark determinant exactly by using the reduction formula for the Wilson fermion. 
Here, we briefly summarize the formula.
For details, see~\cite{Borici:2004bq,Nagata:2010xi,Alexandru:2010yb}. 
For the reduction formula for staggered fermions, 
see~\cite{Gibbs:1986hi,Hasenfratz:1991ax}.

For the preparation of the reduction formula, we define block matrices
\begin{align}
\alpha_i &= \alpha^{ab, \mu\nu}(\vec{x}, \vec{y}, t_i) \nn \\
         &= c_- B^{ab, \mu\sigma}(\vec{x}, \vec{y}, t_i) \; r_{-}^{\sigma\nu} 
         -2  c_+  \kappa \; r_{+}^{\mu\nu} \delta^{ab} \delta(\vec{x}-\vec{y}), 
\\
\beta_i &= \beta^{ab,\mu\nu} (\vec{x}, \vec{y}, t_i), \nn \\ 
        &= c_+ B^{ac,\mu\sigma}(\vec{x}, \vec{y}, t_i)\; r_{+}^{\sigma\nu} 
U_4^{cb}(\vec{y}, t_i) -2 c_- \kappa \; r_{-}^{\mu\nu} \delta(\vec{x}-\vec{y}) 
U_4^{ab}(\vec{y}, t_i).
\end{align}
$c_{\pm}$ are arbitrary scalar except for zero.
$r_\pm = (r \pm \gamma_4)/2$ with the Wilson parameter $r$, where 
the reduction formula can be applied for arbitrary $r$. 
$B$ is the Wilson fermion matrix without the temporal hopping terms. 
$\alpha_i$ describes a spatial hopping at $t_i$, 
while $\beta_i$ describes a spatial hopping at $t_i$ and 
a temporal hopping to the next time slice. They are independent of $\mu$. 

Using the block matrices, the reduction formula is given by 
\begin{subequations}
\begin{align}
\det \Delta(\mu) & = (c_+ c_- )^{-N/2} \xi^{-\Nred/2}  C_0 \det\left( \xi +  Q \right), 
\label{May1010eq2}
\end{align}
with 
\begin{align}
Q   &= (\alpha_1^{-1} \beta_1) \cdots (\alpha_{N_t}^{-1} \beta_{N_t}), 
\label{Eq:2012Jan01eq3}\\
C_0 &= \left(\prod_{i = 1}^{N_t} \det(\alpha_i ) \right),
\label{Eq:2012Jan01eq4}
\end{align}
\end{subequations}
where $\xi=\exp(-\mu/T)$, $N=4N_c N_s^3 N_t$ and $\Nred = N/N_t$. 
The rank of $\alpha_i$ and $Q$ is given by $\Nred$, 
which is reduced to $1/N_t$ compared to the rank of $\Delta$. 
Furthermore, $Q$ and $C_0$ are independent of $\mu$, and the chemical potential 
is separated from the link variables.

The matrix $Q$ describes propagations of quarks from the initial to final time 
slices~\cite{Nagata:2010xi}, and is interpreted as a transfer matrix~\cite{Borici:2004bq,Alexandru:2010yb}. 
Note that all the elements of $Q$ uniformly contain $N_t$ hopping terms in temporal direction, 
which enables to separate $\mu$ from $Q$.  
$C_0$ consists of the closed loops without temporal hopping. 
Then, $C_0$ is also independent of  $\mu$.

To obtain $\det \Delta$, we need to evaluate $\det (Q+\xi)$. 
Here we calculate the eigenvalues $\lambda$ for $|Q-\lambda I|=0$. 
Although the eigen problem requires large numerical cost, there is an advantage. 
Once we obtain $\lambda$, the quark determinant is the analytic function 
of $\mu$. Then, the value of $\det \Delta(\mu)$ is obtained 
for arbitrary $\mu$, which is useful for MPR.  
Other methods such as LU decomposition of $Q+\xi$ can be used instead of solving the eigenvalue problem for $Q$. 
In this case, we need to perform the LU decomposition for each $\mu$. 

With the eigenvalues of $Q$, we obtain
\begin{subequations}
\begin{align}
\det \Delta (\mu) &=  C_0  \xi^{-\Nred/2}\prod_{n=1}^{\Nred} (\lambda_n + \xi), 
\label{Nov292011eq1}\\
              &= C_0 \sum_{n=0}^{\Nred}c_n \xi^{n-\Nred/2}
              = C_0 \sum_{n=-\Nred/2}^{\Nred} c_n \xi^n,
\label{Jan1511eq1} 
\end{align}
\label{Jan1511eq3}%
\end{subequations}%
where we set $c_\pm =1$ for simplicity. 
Here we describe the determinant in two expressions:  a product form Eq.~(\ref{Nov292011eq1}), 
and a summation form Eq.~(\ref{Jan1511eq1}). The second one denotes the fugacity 
expansion of the quark determinant, where fugacity coefficients $c_n$ 
are polynomials of the eigenvalues $\lambda_n$~\cite{Nagata:2010xi}.

\subsection{Taylor expansion of EoS}
\label{subsec2d}
Next we consider the Taylor expansion for the EoS. 
A noise method is often used to calculate Taylor coefficients.
In this work, however, the derivatives are exactly obtained even for higher order terms 
by using the reduction formula, which excludes errors 
caused by noise methods. 
As we will explain below, the thermodynamical quantities are 
obtained by using the eigenvalues of the reduced matrix both 
in the Taylor expansion and MPR methods. 
This provides an equal-footing basis for the comparison and consistency of the two methods. 

The deviation of the pressure is expanded in powers of $\mu/T$ at $\mu=0$ 
as follows
\begin{subequations}
\begin{align}
\frac{\delta p(\mu,T) }{T^4} & = \sum_{n=2,4,\cdots}^\infty c_n(T) \left(\frac{\mu}{T}\right)^n_,
\end{align}%
where $c_n$ are Taylor coefficients at $\mu=0$ given by 
\begin{align}
c_n = \left. \frac{1}{n!} \left(\frac{N_t}{N_s}\right)^3 
T^n \frac{\del^n \ln Z_{GC}}{\del \mu^n} \right|_{\mu\to 0  \; .}
\label{Eq:Dec0310no1}
\end{align}
\end{subequations}
The number density and number susceptibility are given by 
\begin{subequations}
\begin{align}
\frac{ n}{T^3} & = \sum_{m=2,4,\cdots}^\infty m \cdot c_m(T) \left(\frac{\mu}{T}\right)^{m-1}_, \\
\frac{\chi}{T^2} & = \sum_{n=2,4,\cdots}^\infty n(n-1)\cdot c_n(T) \left(\frac{\mu}{T}\right)^{n-2}_ .
\end{align}%
\label{Eq:2012Mar20eq1}%
\end{subequations}%
The $n$-th derivative of the grand partition function $Z_{GC}^{(n)}= (T\del/\del\mu)^n Z_{GC}$ is given by
\begin{align}
\frac{Z_{GC}^{(n)} }{Z_{GC}} &= \left\bra \frac{(T \del/\del \mu)^n [\det \Delta(\mu)]^{N_f} }{[\det \Delta(\mu)]^{N_f}} \right\ket. 
\label{Jan2511eq8}
\end{align}
Derivatives of $\det \Delta$ are obtained from Eq.~(\ref{Nov292011eq1}) and Eq.~(\ref{Jan1511eq1}).
Equation (\ref{Jan1511eq1}) gives
\begin{align}
T^k \frac{\del^k}{\del \mu^k} \det \Delta (\mu) &= C_0 \sum_{n=0}^{\Nred} 
(\Nred/2-n)^k c_n \xi^{n-\Nred/2}, 
\label{Jan2711eq5}
\end{align}
which holds for arbitrary $k$. 
To derive derivatives of the product form Eq.~(\ref{Nov292011eq1}), 
we rewrite it as
\begin{align}
\det \Delta (\mu) & = \exp\left(\log(C_0  \xi^{-\Nred/2}\prod_{n=1}^{\Nred} (\lambda_n + \xi)) \right).
\label{Jan1511eq2}
\end{align}
Then, derivatives are straightforwardly obtained  by algebraic calculations. 
We use the product form, because it is easier to calculate than the 
summation form. The summation form is used for the check. 

\section{Result}
\label{sec3}

\subsection{Simulation setup}
\label{subsec3a}
We consider the clover-improved Wilson fermions with $N_f=2$ and 
RG-improved gauge action. 
Simulations were performed mostly on a $N_s^3\times N_t = 8^3\times 4$ 
lattice. 
We considered $29$ values of $\beta$ in the interval $1.5\le \beta \le 2.4$ for $N_s=8$. 
Simulations on a $10^3\times 4$ lattice were also performed for near $\beta_{\rm pc}$ to 
investigate the finite size effect.  
We considered $16$ values in the interval $1.8\le \beta \le 1.95$ for $N_s=10$. 
The value of the hopping parameter $\kappa$ was determined for each $\beta$ 
by following the line of the constant physics with $m_{\rm PS}/m_{\rm V}=0.8$ in Ref.~\cite{Ejiri:2009hq}. 
The clover coefficient $\Csw$ was determined by using a result obtained 
in the one-loop perturbation theory : $\Csw = ( 1- 0.8412 \beta^{-1})^{-3/4}$. 

Gauge configurations were generated at $\mu=0$ with the hybrid Monte Carlo simulations.
The setup for the molecular dynamics was as follows: a step size $\delta \tau = 0.02$, 
number of the step $N_{\tau}=50$ and length  $N_{\tau}\delta\tau =1$. 
The acceptance ratio was more than 90 \% for $N_s=8$ and 80 \% for $N_s=10$. 
HMC simulations were carried out for 11, 000 trajectories for each parameter set. 
For all the ensemble, the first 3,000 trajectories were removed as thermalization. 
The eigenvalues of the reduce matrix $Q$ were calculated for each 20 HMC steps,  
and 400 sets of the eigenvalues were collected for each ensemble. 

We show the estimation of computation time for the reduction 
formula, where we consider the following three steps ; 
calculation of the overall factor $C_0$ (\ref{Eq:2012Jan01eq4}), the construction of the 
matrix $Q$ (\ref{Eq:2012Jan01eq3}) which includes the inverse matrices, the solving the eigenvalue problem. 
The details of the numerical procedure are as follows.
LAPACK Library ZGETRF was used for the LU factorization of $\alpha_i$ and 
the calculation of $C_0$. 
ZGETRI together with ZGETRF were used to obtain the inverse of $\alpha_i$, 
and ZGEMM in BLAS was used, then $Q$ was constructed. 
ZGESS in LAPACK was used to obtain eigenvalues of $Q$. 
NEC SX-9 at Osaka University was used in the calculations. 
Taking the average over 400 configurations, we evaluate the total time 
for these three procedure,  
and then further we take the average over some parameter sets. 
Estimated time was $750$ sec for $8^3\times 4$ and $4 000$ sec 
for $10^3\times 4$. They are not scaled by $V^3$, probably 
due to overhead time to construct $Q$. 
As a basis for comparison, we evaluated CPU time for 1000 HMC trajectories
 with the molecular dynamics setup explained above, where the standard 
CG algorithm was used.  
We spent about 61 000 sec for $8^3\times 4$, and 112 000 sec for $10^3\times 4$
in average. 
As a benchmark, the ratio (Time for 400 reduction) / (Time for 10, 000 HMC) 
is 
\begin{align}
\frac{750 \times 400}{61 000\times 10} &= 0.5   \;\; ( 8^3\times 4) \nn, \\
\frac{4 000 \times 400}{112 000 \times 10 } &= 1.4 \;\; (10^3\times 4)  \nn. 
\end{align}
The numerical cost of the reduction formula was  almost the same order as 
that of 10, 000 HMC update in $8^3\times 4$ or $10^3\times 4$ lattice in the present 
calculation setup.
If one performs the determinant calculation of the original Wilson 
matrix, the above quantity would become about $N_t^2= 16$ times larger.

\subsection{Fluctuation of the quark determinant}
\label{subsec3b}
First, we investigate the fluctuation of the quark determinant. 
\begin{figure}
\includegraphics[width=7cm]{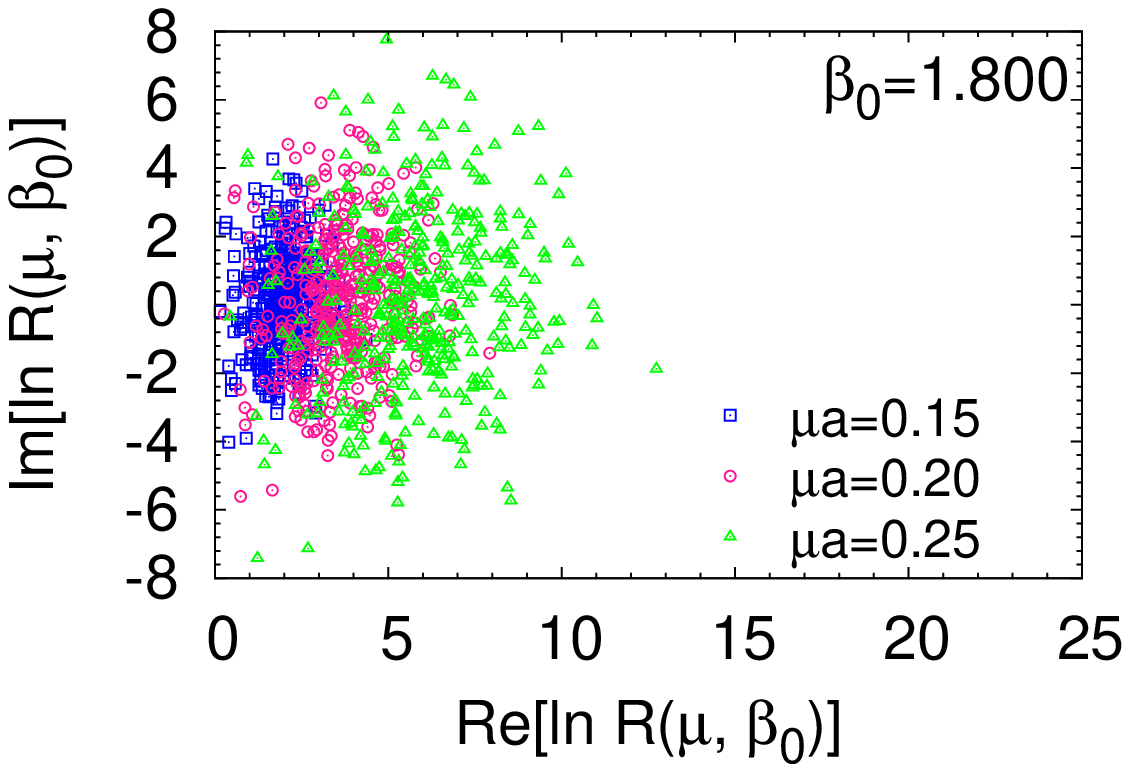}
\includegraphics[width=7cm]{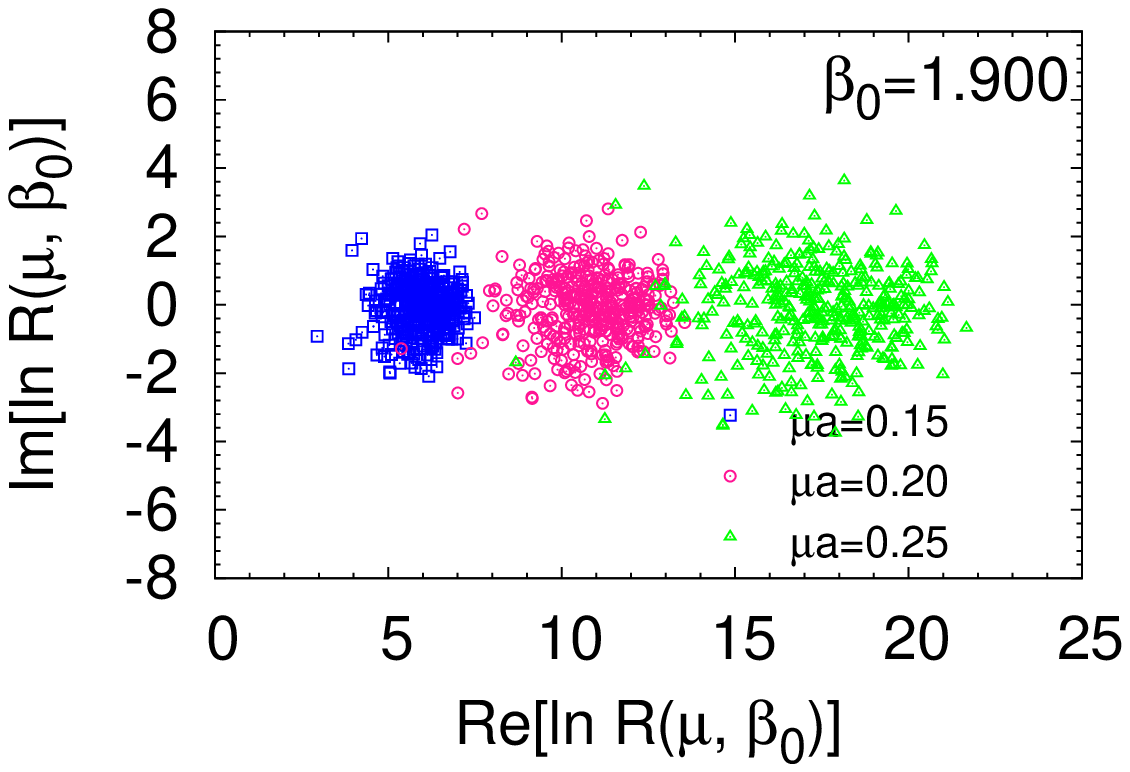}
\caption{The scatter plot of the quark determinant on the complex plane. 
}
\label{Oct142011Fig1}
\end{figure}
Figure~\ref{Oct142011Fig1} shows the scatter plot of 
$N_f\ln \det \Delta(\mu)/\det \Delta(0)=\ln R(\mu, \beta_0)_{(0,\beta_0)}$.
We show the results for $\beta_0=1.8 (T/\Tpc\sim 0.93)$ and $1.9 (1.08)$.
The quark determinant shows different $\mu$-dependence corresponding to the value of $\beta_0$. 
It increases mainly in magnitude at $\beta_0=1.9$ (high $T$), 
while it increases in phase at $\beta_0=1.8$ (low $T$). 
Near $\betapc(\sim 1.86)$, the quark determinant fluctuates between 
low-$T$ and high-$T$ states. 

\begin{figure}[htbp]
\includegraphics[width=7cm]{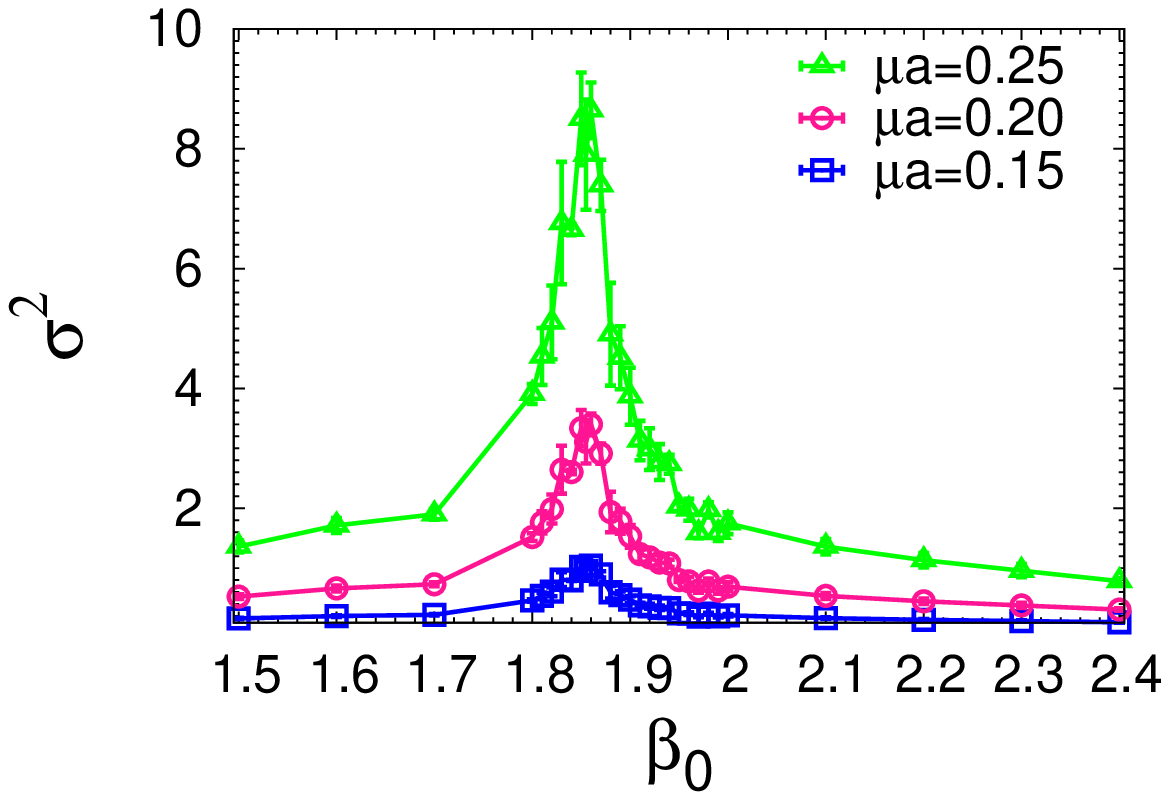}
\includegraphics[width=7cm]{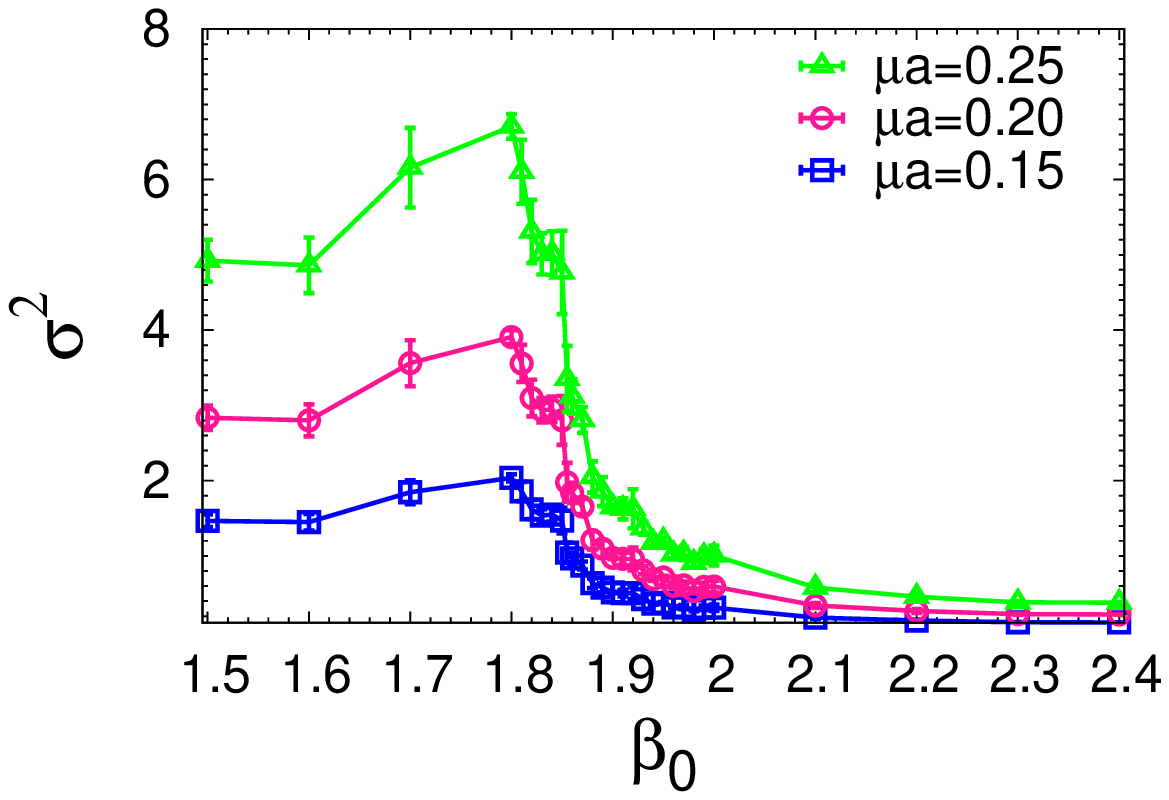}
\caption{The fluctuation of the quark determinant as a function of $\beta_0$. 
$\sigma^2= \frac{1}{n}\sum_i (x_i-\bar{x})^2, \bar{x}=\frac{1}{n}\sum_i x_i$, 
where $x={\rm Re}[\ln R(\mu, \beta_0)_{(0,\beta_0)}]$ for left panel and $x={\rm Im}[\ln R(\mu, \beta_0)_{(0,\beta_0)}]$ 
for right panel.}
\label{QDet2011Oct21Fig1}
\end{figure}
Figure~\ref{QDet2011Oct21Fig1} illustrates the behavior of the fluctuation of 
the quark determinant defined as the standard deviation. 
The real part of $\ln R$, which is the power of $|R|$, 
shows a peak near the crossover transition $\betapc$, which is caused by 
the fluctuation between low- and high-$T$ states. 
The peak causes the contamination of unimportant configurations and implies 
the severe overlap problem. The peak becomes prominent for $\mu a>0.2$.  
Except for the vicinity of $\betapc$, the fluctuation is not so large compared to the present statistics
at least for small $\mu$, and therefore the overlap problem is not so severe.

For the imaginary part of $\ln R$, which is the phase of $R$, the fluctuation is 
large for near and below $\betapc$, and small at large $\beta$. 
It was pointed out ~\cite{Ejiri:2004yw} that the fluctuation of the phase of the 
reweighting factor is not suppressed by the MPR method because the gauge part is real. 
If the phase goes over $\pi/2$, the determinant changes the sign, and causes 
the sign problem.  
Adopting the standard deviation as a criterion, the onset of the problem is 
$\mu a\sim 0.2$ near $\betapc$. This imposes an applicable limit of MPR on 
the $8^3\times 4$ lattice in the present simulation setup. 
We limit our analysis on the thermodynamical quantities up to $\mu a=0.20$. 
Applicable range of MPR in the present work is smaller than that of 
staggered fermions investigated in Ref.\cite{Csikor:2004ik}. 
This difference may be caused by small statistics.

The severity of the problems is roughly classified into three cases according to 
temperatures.  
At high temperatures, the fluctuation is small both for the real and imaginary 
parts, and the sign problem and overlap problem is not severe. 
Near $\betapc$, both the real and imaginary parts fluctuate rapidly. 
At low temperatures, the phase fluctuates rapidly, while the fluctuation 
of the real part is not so large. 

\subsection{Fluctuation of the reweighting factor and Reweighting line}
\label{subsec3c}
\begin{figure}
\includegraphics[width=5cm]{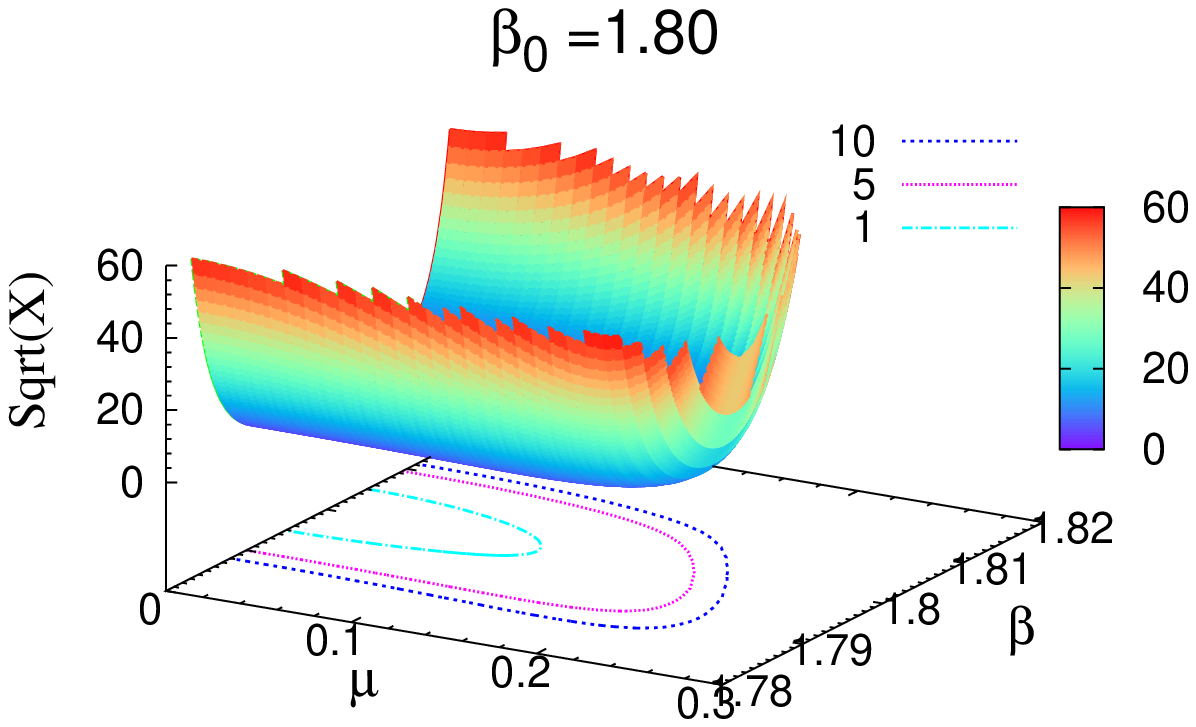}
\includegraphics[width=5cm]{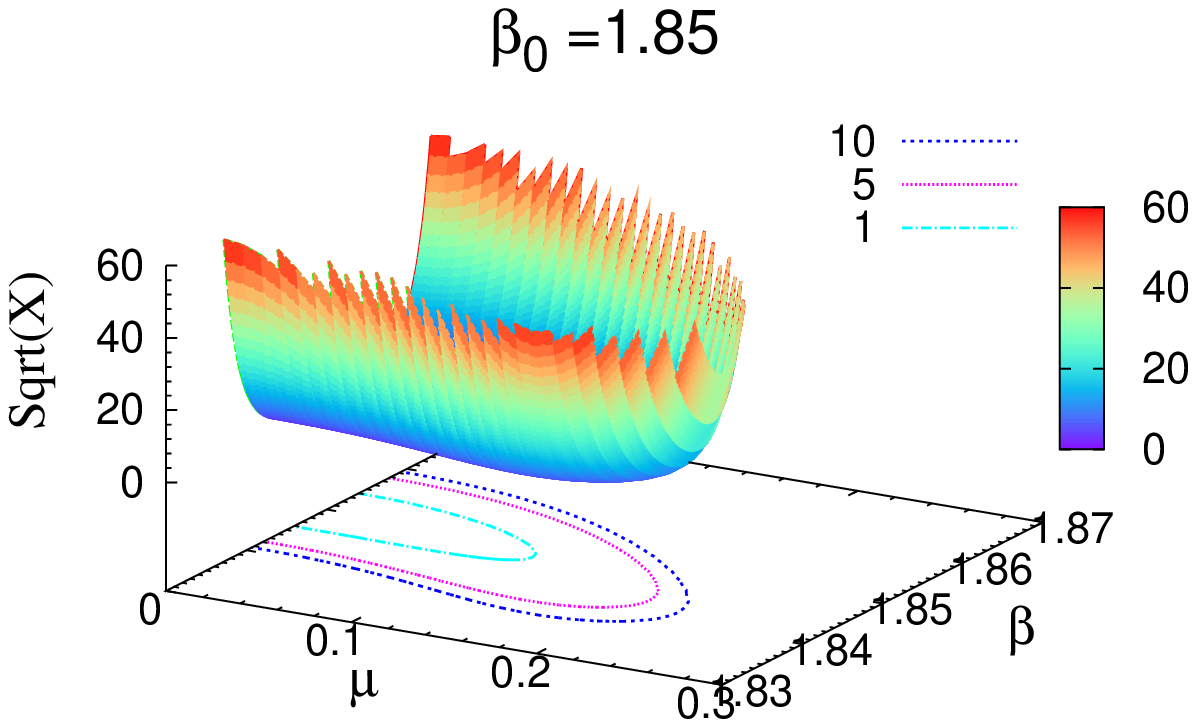}
\includegraphics[width=5cm]{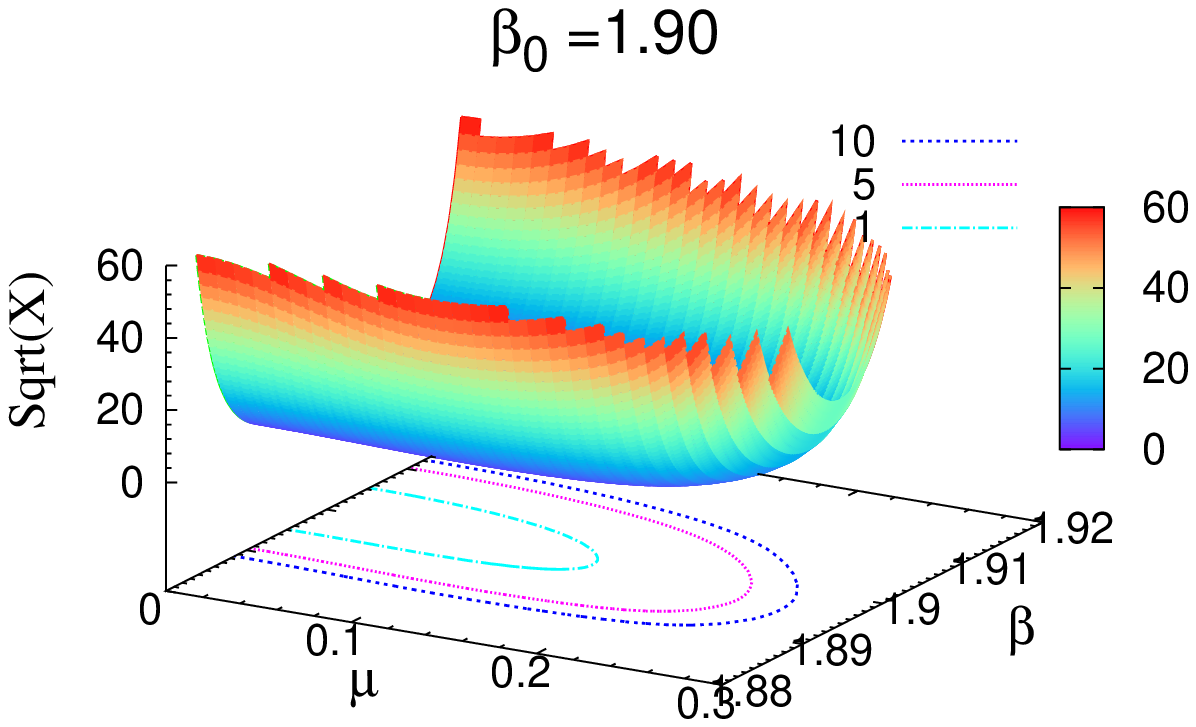}
\includegraphics[width=5cm]{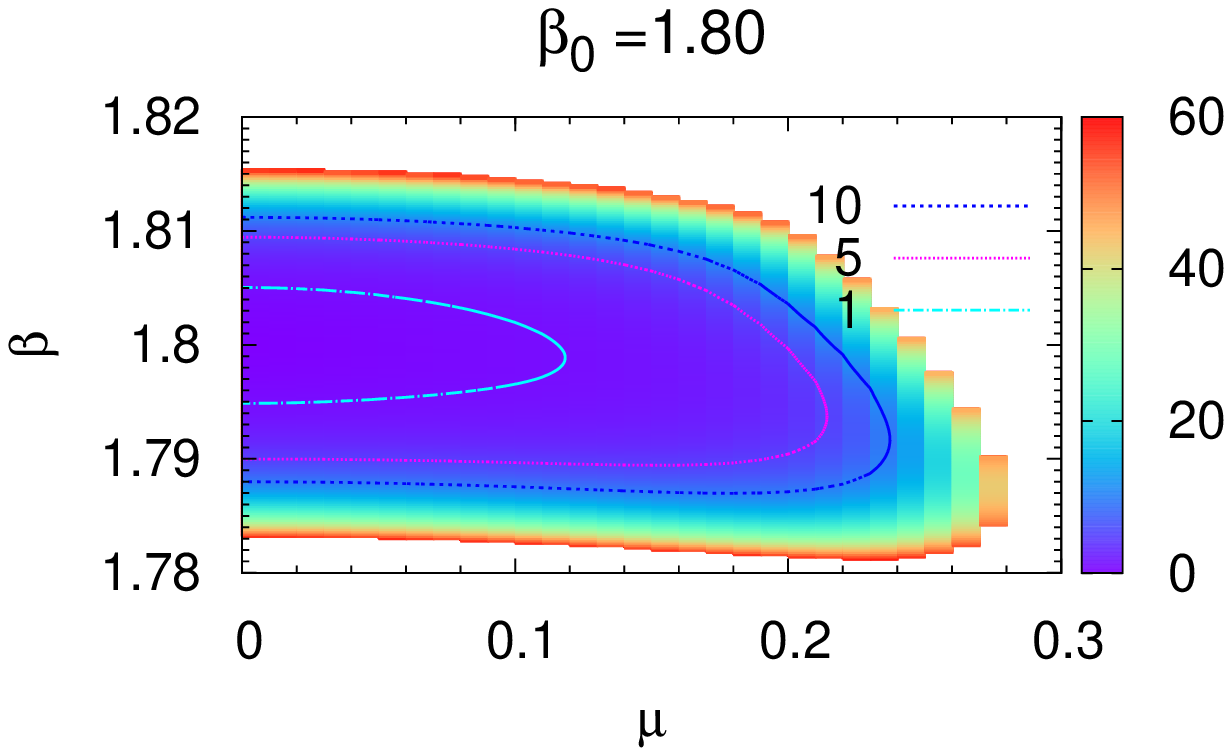}
\includegraphics[width=5cm]{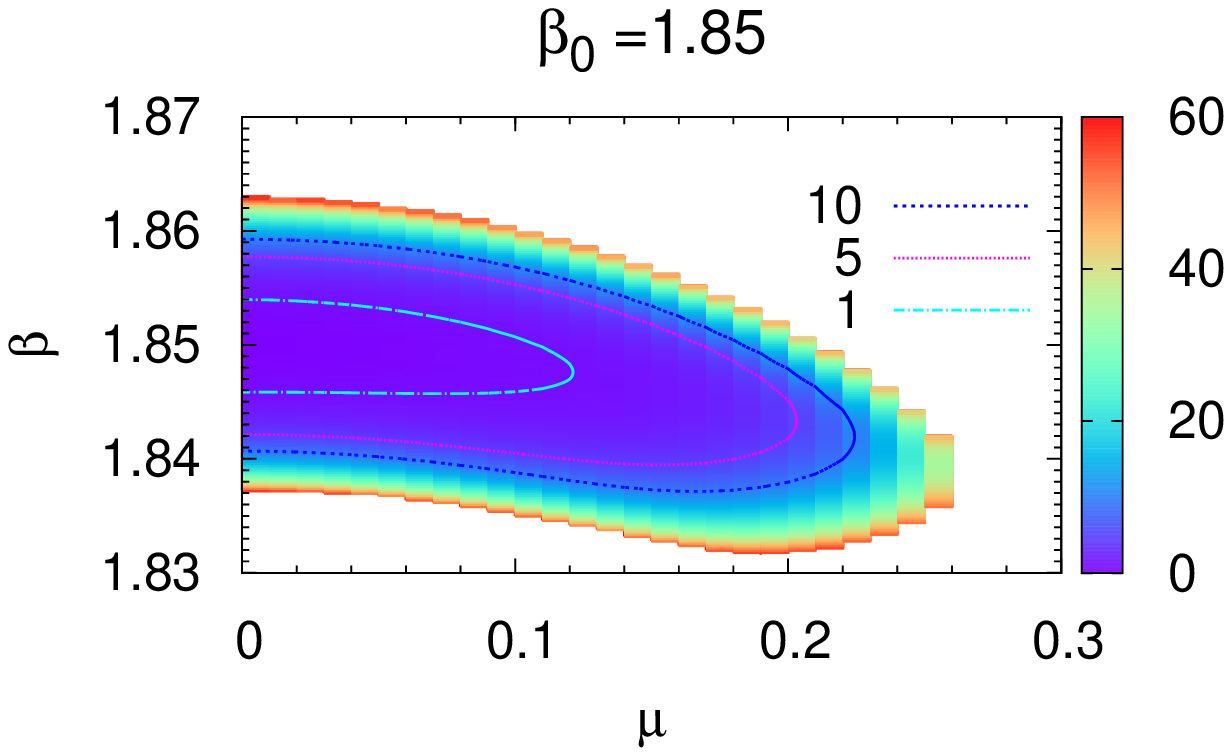}
\includegraphics[width=5cm]{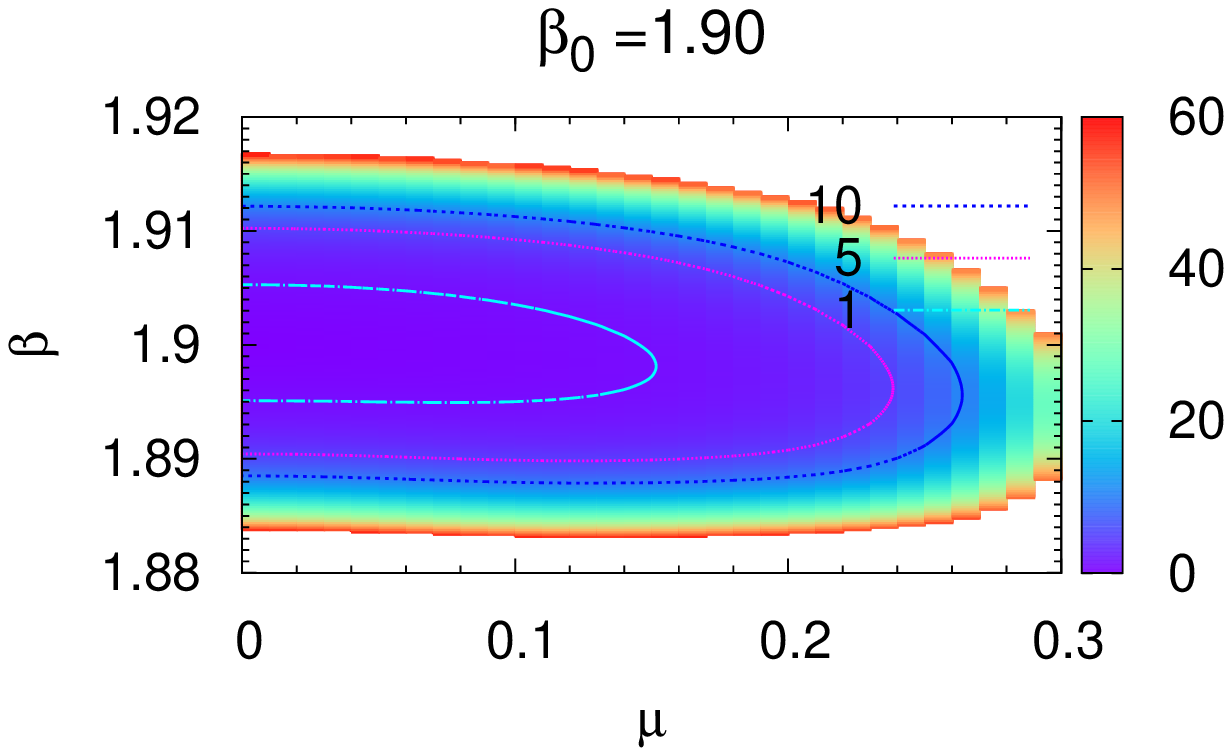}
\caption{
Fluctuation of the total reweighting factor (top panels) and its 
contour plot on the $\mu$-$\beta$ plane (bottom panels).
The simulation points are $\beta_0=1.80, 1.85$ and $1.90$ for
left, middle and right panels respectively. Here we take the 
absolute values of the fluctuation $X=\bra | R - \bra R\ket_0 |^2 \ket_0$. 
}
\label{Sep252011Fig2}
\end{figure}
\begin{figure}
\includegraphics[width=7cm]{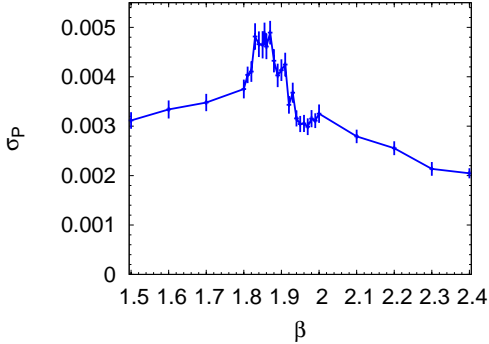}
\caption{Variation of the plaquette distribution extracted from  
the Gaussian fit $f(P)= \exp(-(P-\bra P\ket)^2/(2\sigma_P^2))$ with 
$P$ being the plaquette. }
\label{Oct272011Fig1}
\end{figure}
Next, we consider the fluctuation of the reweighting factor $X$. Here 
we modify the condition to $X = \bra | R - \bra R\ketSP |^2 \ketSP$. It is 
an alternative choice to take the real part of $R$. In the calculation of thermodynamical
quantities, we limit our study to the region where the fluctuation
of the phase is small. Then, we can use either of the absolute value or real part.   

The contour plot of $X$ in Fig.~\ref{Sep252011Fig2} illustrates how $X$
increases in the shift of the parameters from simulation points. 
The shape of $X$ is related to the distribution of the quark determinant and of the plaquette, 
see Figs.~\ref{QDet2011Oct21Fig1}, \ref{Sep252011Fig2} and \ref{Oct272011Fig1}. 
The rapid fluctuation of the plaquette makes the valley steep in $\beta$ direction, 
while that of the quark determinant makes the valley steep in $\mu$ direction. 

Near $\betapc$, both the plaquette and the quark determinant fluctuate rapidly, 
which makes the valley steep and results in narrowing the small fluctuation domain. 
For this case, the valley curves downward, and $X$ remains small due to the 
cancel of the contributions of the plaquette and quark determinant.

To avoid the overlap problem, the fluctuation $X$ needs to be suppressed. 
The reweighting line is taken along the valley of $X$ for each ensemble. 

\subsection{Consistency of MPR and Taylor expansion for EoS}
\label{subsec3d}
\begin{figure}
\includegraphics[width=7cm]{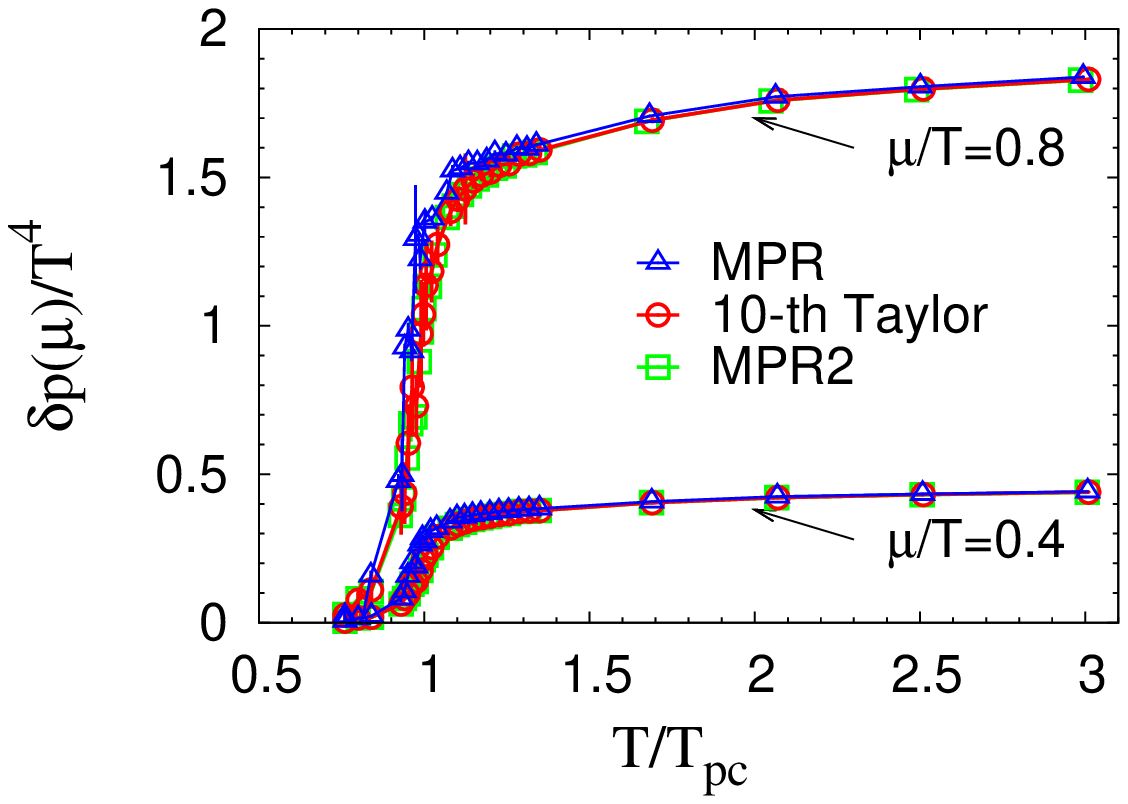}
\includegraphics[width=7cm]{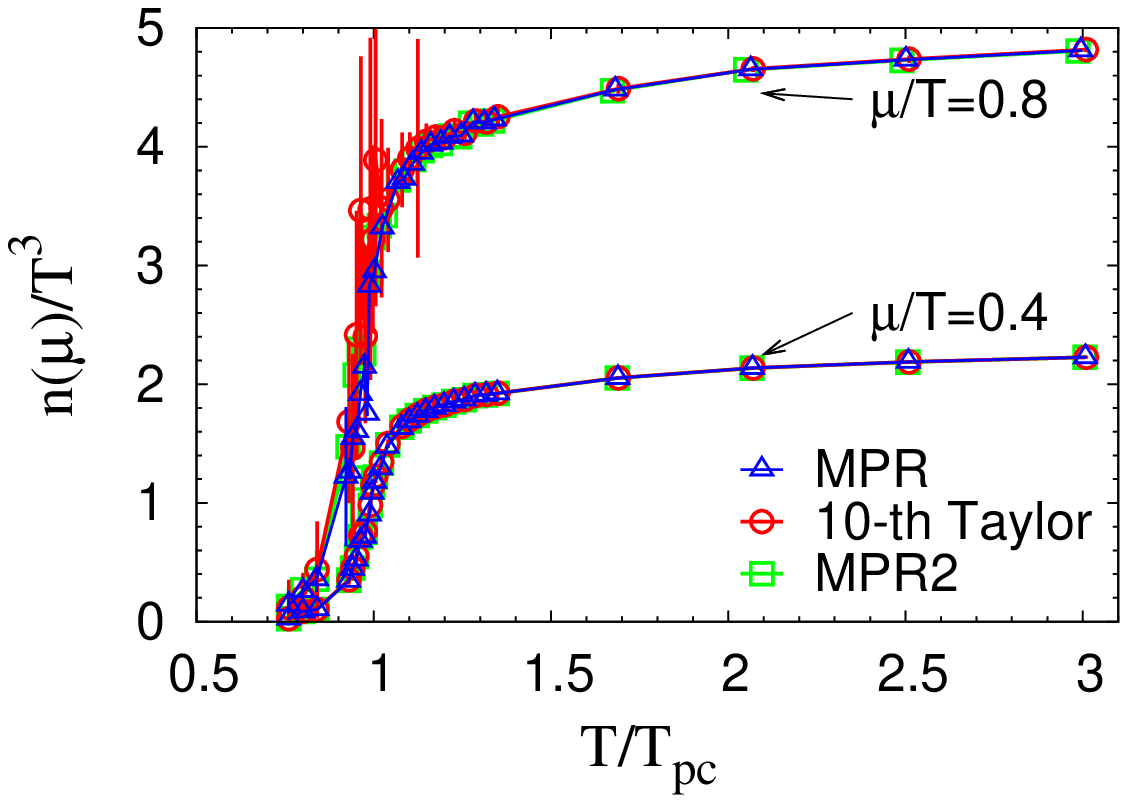}
\includegraphics[width=7cm]{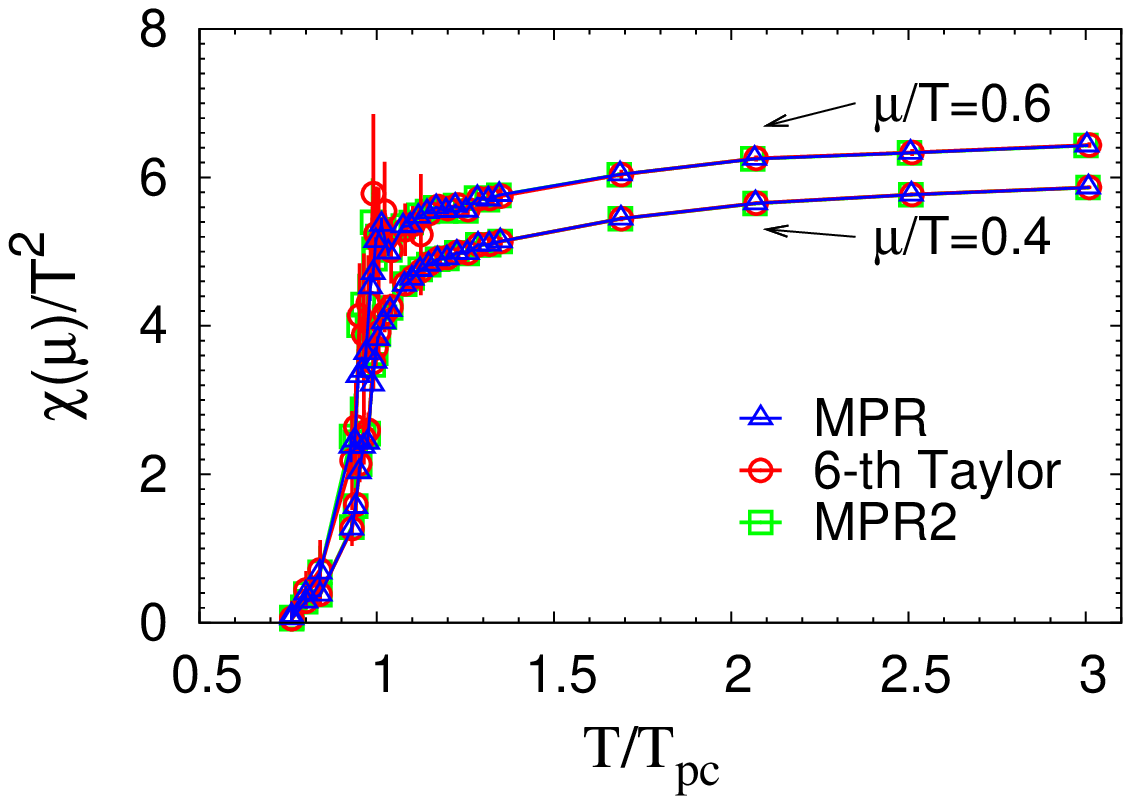}
\caption{Thermodynamical quantities obtained from 
Taylor expansion (circle),  MPR with the fluctuation 
minimum condition (triangle) and MPR with Eq.~(\ref{Nov302011eq1}) (square). 
}
\label{Sep262011Fig1}
\end{figure}
\begin{figure}[htbp]
\includegraphics[width=5cm]{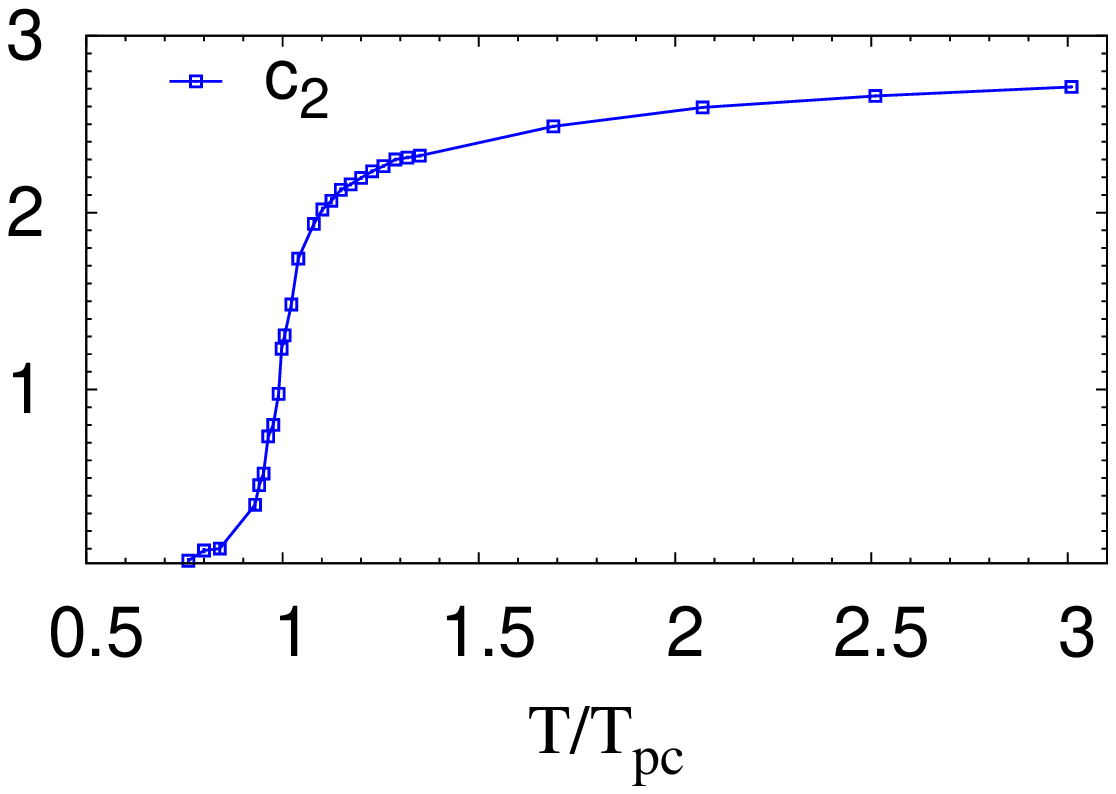}
\includegraphics[width=5cm]{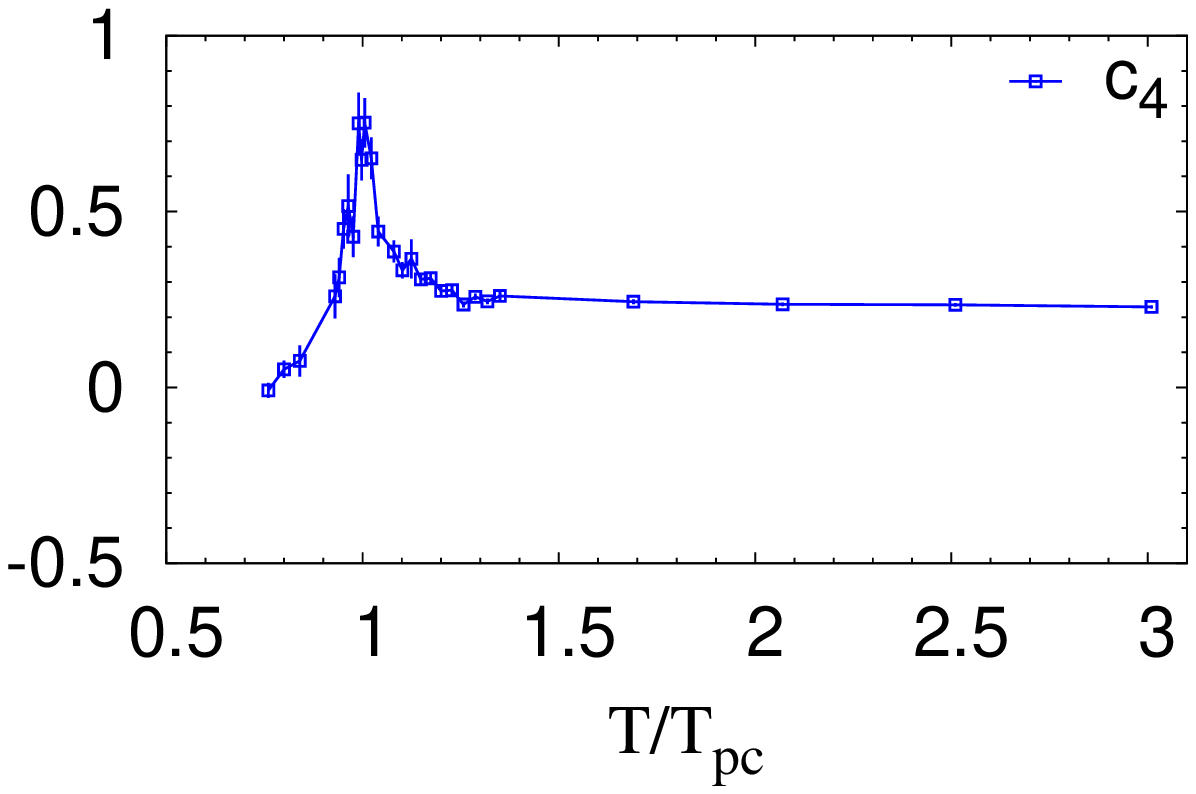}
\includegraphics[width=5cm]{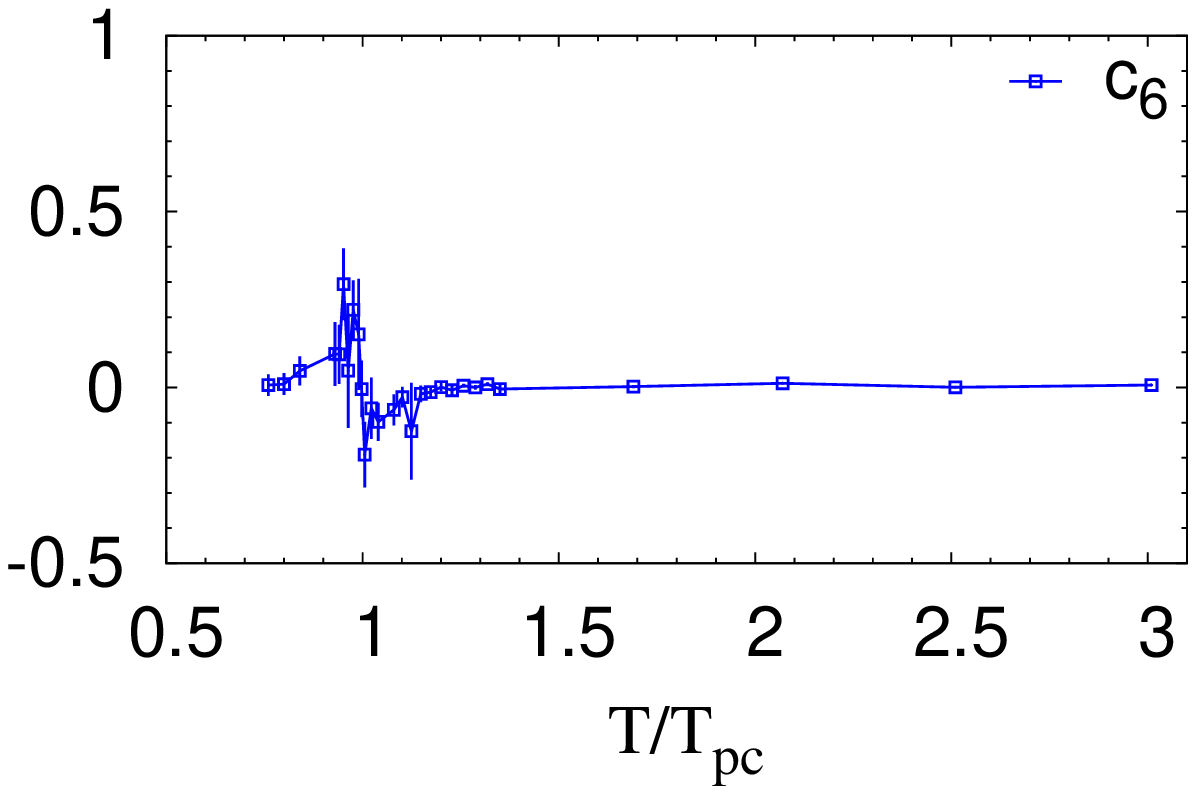}
\includegraphics[width=5cm]{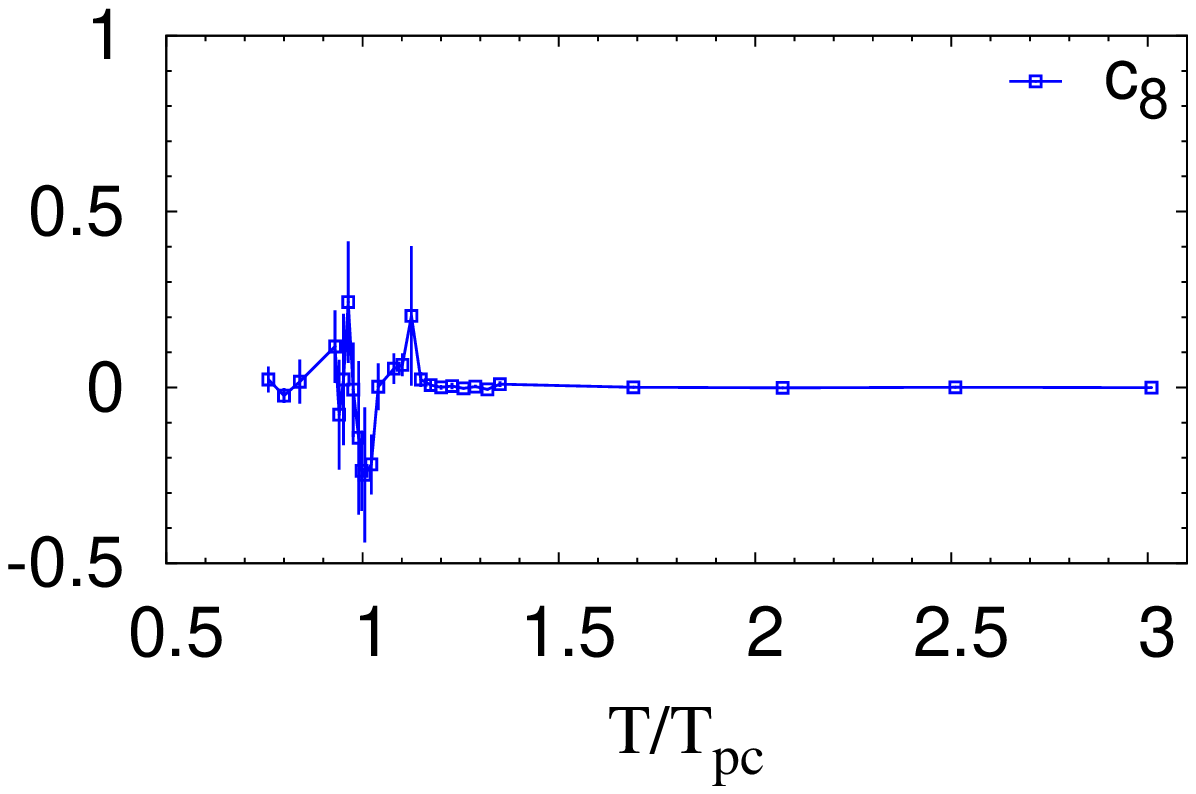}
\includegraphics[width=5cm]{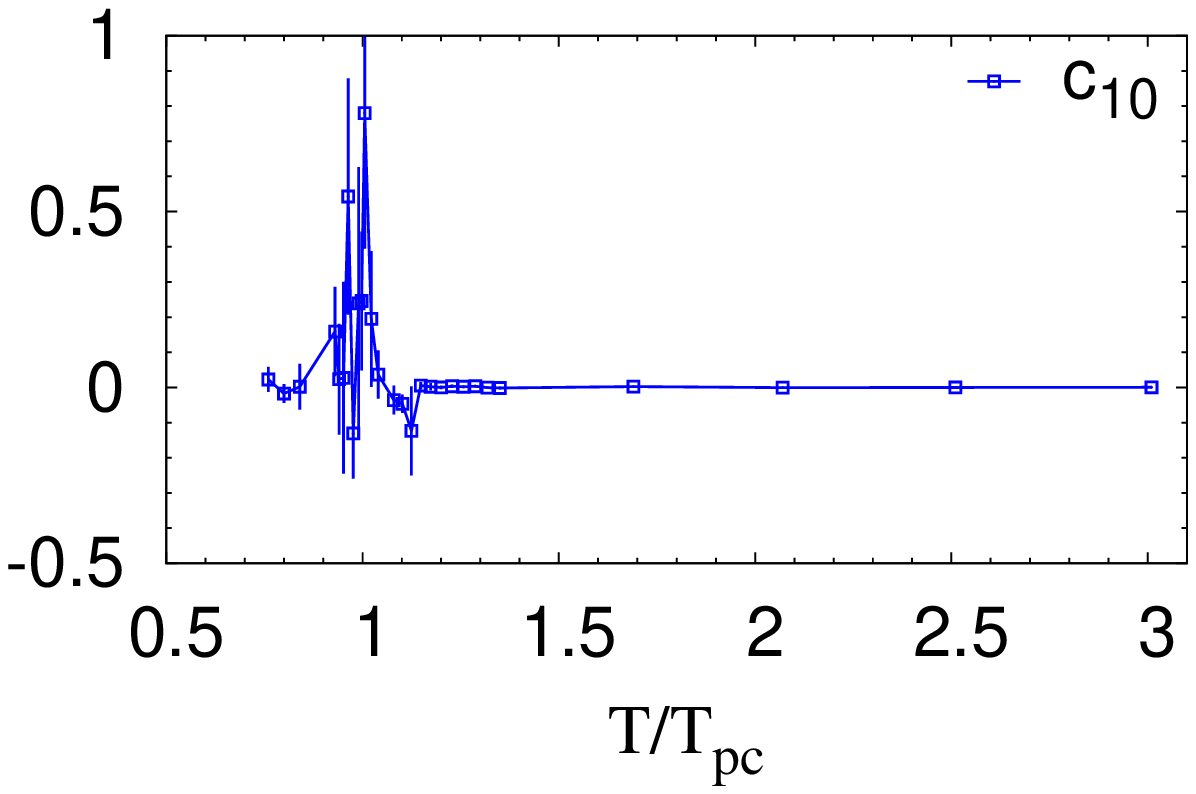}
\caption{Taylor coefficients $c_n, (n=2,\cdots 10)$.}
\label{Sep232011Fig1}
\end{figure}

Thermodynamical quantities are shown in Fig.~\ref{Sep262011Fig1}. 
To obtain the values of $T$ from $\beta$, we use the data in Ref.~\cite{Ejiri:2009hq}.
The EoS and number density for the Taylor expansion contains up to tenth order, 
while the susceptibility up to sixth order. The Taylor coefficients 
given in Eq.~(\ref{Eq:Dec0310no1}) are shown in Fig.~\ref{Sep232011Fig1}.
MPR and Taylor expansion methods are almost consistent up to $\mu/T\sim 0.8$ for the EoS and quark number density. 
For the susceptibility, the consistency holds for up to $\mu/T\sim 0.6$, 
while errors become large for $\mu/T> 0.6$ particularly near $\Tpc$.

Figure~\ref{Sep262011Fig1} also shows the results obtained from the 
equation of the reweighting line given in Eq.~(\ref{Nov302011eq1}).
It turns out that the equation of the reweighting line is almost consistent with 
the fluctuation minimum condition.
Next, we see the Taylor coefficients. 
$c_2$ and $c_4$ are consistent with those obtained by the same action 
and larger lattice $16^3 \times 4$~\cite{Ejiri:2009hq} probably due to the crossover nature of the transition at small $\mu$. 
The Taylor series converges up to ${\cal O}((\mu/T)^4)$ for high $T$, which is
consistent with the expected behavior from free-quark-gluon picture. 
On the other hand, the convergence is slow near and below $\Tpc$. 
$c_n$ oscillates and the number of the oscillation increases with $n$. 
This behavior was observed in a  Polyakov-quark-meson model with $2+1$ 
flavors~\cite{Schaefer:2009st}.
Statistical errors become larger for higher coefficients. 
The inclusion of $c_8$ and $c_{10}$ causes large errors for the number 
susceptibility $\chi$. 
The errors become significant for large $\mu/T\sim 1$ for EoS and $\mu/T\sim 0.8$ for $\chi$.  

\begin{figure}[htbp]
\begin{center}
\includegraphics[width=7cm]{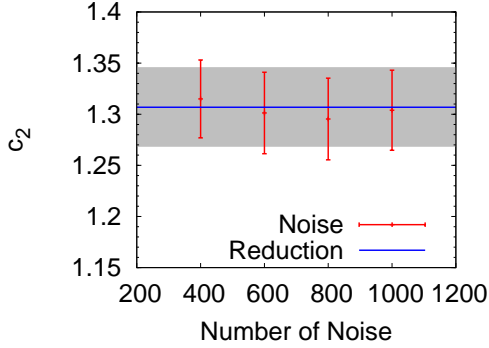}
\caption{The Taylor coefficient $c_2$ at $T/\Tpc=1(\beta=1.86)$ 
obtained from the noise method with different number of noise vectors. 
The horizontal line is the value obtained from the reduction formula, 
where the errorbar is denoted by the gray region.}
\label{Mar202012Fig1}
\end{center}
\end{figure}
The comparison with a noise method for $c_2$ is shown in 
Fig.~\ref{Mar202012Fig1}, where the trace of an operator $A$ is calculated 
by $\mbox{tr} A = (N_r^{-1}) \sum_{i=1}^{N_r} (v^{(i)})^* A v^{(i)}$, 
where $v^{(i)}, (i=1,2,\cdots, N_r)$ is noise vectors and 
$N_r$ is the number of the noise vectors. 
We employ the noise vectors for all the indices, i.e., the color, Dirac
and coordinate space, 
$N_r^{-1} \sum_{i=1}^{N_r} v_{a,\alpha,\vec{x}}^{(i)} 
(v_{b,\beta,\vec{y}}^{(i)})^* 
= \delta_{a,b} \delta_{\alpha,\beta} \delta_{\vec{x},\vec{y}}$. 
It turns out that 400 noise vectors are almost enough for the noise method to 
reproduce $c_2$ of the reduction formula both in the average value and 
errorbar. 
Note that the number of the noise may be reduced further by the 
improvement of the noise methods.
For each measurement, the noise method slowly converges according to 
$O(1/\sqrt{N_r})$, and large number of noise vectors are needed 
to reproduce the result of the reduction formula.  
Taking the ensemble average improves the convergence, which 
allows to use fewer number of the noise vectors. 
The computational time for one measurement of $c_2$ with 
BiCGStab algorithm was about 240, 320 and 400 sec for $N_r=$ 600, 800 and 1000, respectively, while 
the time for the reduction formula was about 1000 sec. 
For $c_2$, the noise method is several times faster than the reduction formula. On the other hand, the reduction formula provides higher order 
coefficients with small additional calculation. 
For higher-order Taylor coefficients, the reduction formula becomes 
faster than the noise method.
However, it should be noted that the reduction formula is limited to 
small lattice size.

Here we comment on the difference of the errorbars,  and 
on the applicable limit of the two approaches. 
In our approach, the Taylor expansion and MPR methods are obtained 
from the same quantities, i.e., the eigenvalues of the reduced matrix. 
The thermodynamical quantities are defined in Eq.~(\ref{Nov092011eq1}) for 
the MPR method and in Eq.~(\ref{Eq:2012Mar20eq1}) for the Taylor expansion 
method. 
In the Taylor expansion, the numerical errors of the thermodynamical 
quantities mainly come from higher-order Taylor coefficients, see Fig.~\ref{Sep232011Fig1}.
The derivatives of the pressure, $n/T^3$ and $\chi/T^2$, are sensitive to 
higher $c_n$ because of the multiplicative factors in Eq.~(\ref{Eq:2012Mar20eq1}).
For instance, the tenth term $c_{10}$ is enhanced by the factor $10$ and 
$10\times 9$ in $n/T^3$ and $\chi/T^2$, respectively. 
This is the origin of the large errors in Fig.~\ref{Sep262011Fig1} 
and restricts the applicable limit of the Taylor expansion. 
For large $\mu/T$, higher-order coefficients become important, 
and as a consequence, the Taylor expansion of the EoS is breakdown, 
which happens at $\mu/T\sim 0.8$ near $T\sim \Tpc$ for $\delta p/T^4$.  

In the MPR method, the numerical errors come from 
statistical fluctuation of the reweighting factor and observables. 
The MPR requires only the first and second derivative terms of the fermion 
determinant in the calculation of $n/T^3$ and $\chi/T^2$, 
see Eq.~(\ref{Nov092011eq1}). 
The fluctuation of the reweighing factor is suppressed if the 
parameters change along the small fluctuation region, see 
Fig.~\ref{Sep252011Fig2}. 
The major origin of the difference in the errorbars is the 
calculation of higher-order derivative terms, which is contained only 
in the Taylor expansion method and not in the MPR method~\ref{Sep232011Fig1}.
As shown in Fig.~\ref{QDet2011Oct21Fig1}, the fluctuation of the 
imaginary part of the reweighting factor becomes large about 
$\mu a \sim 0.2 (\mu/T\sim 0.8)$ near $T\sim \Tpc$, which is also near 
the edge of the small fluctuation domain in Fig.~\ref{Sep252011Fig2}. 
Thus, the applicable range of the two methods are consistent, although the 
numerical errors appear in different way. 
This is natural consequence of that the fermion determinant and its derivatives are 
given by the same quantities $\lambda_n$ of the reduced matrix, 
hence their fluctuations are correlated.

Thus, the MPR and Taylor expansion methods suffer from the different 
difficulties. Hence, their consistency implies that 
the truncation error of the Taylor expansion method and overlap problem of 
MPR are not serious and that the obtained thermodynamical quantities are 
reliable in these regions in spite of these difficulties. 

Note that the fluctuation of imaginary part of the reweighting 
factor depends on the lattice volume, and the applicable range 
of the reweighting becomes smaller as the lattice volume increases. 
We will discuss this point later. 

\subsection{Consistency with imaginary chemical potential approach}
\label{subsec3e}
\begin{figure}
\includegraphics[width=7cm]{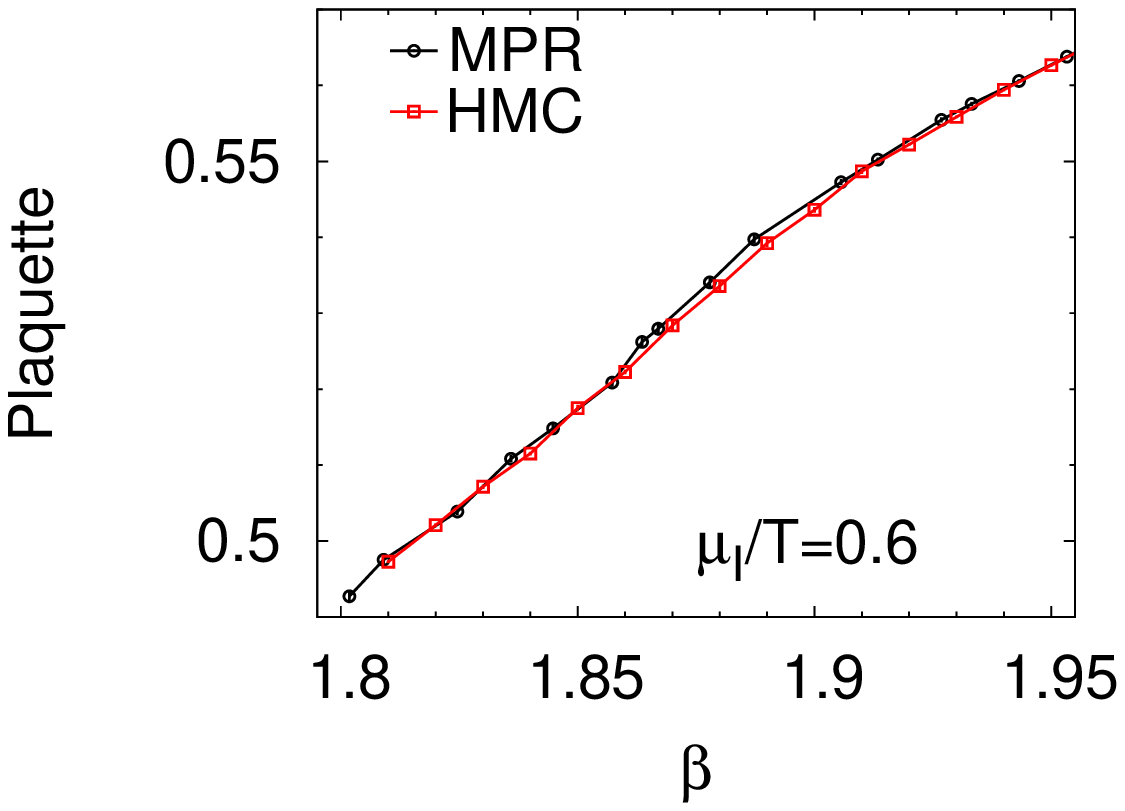}
\includegraphics[width=7cm]{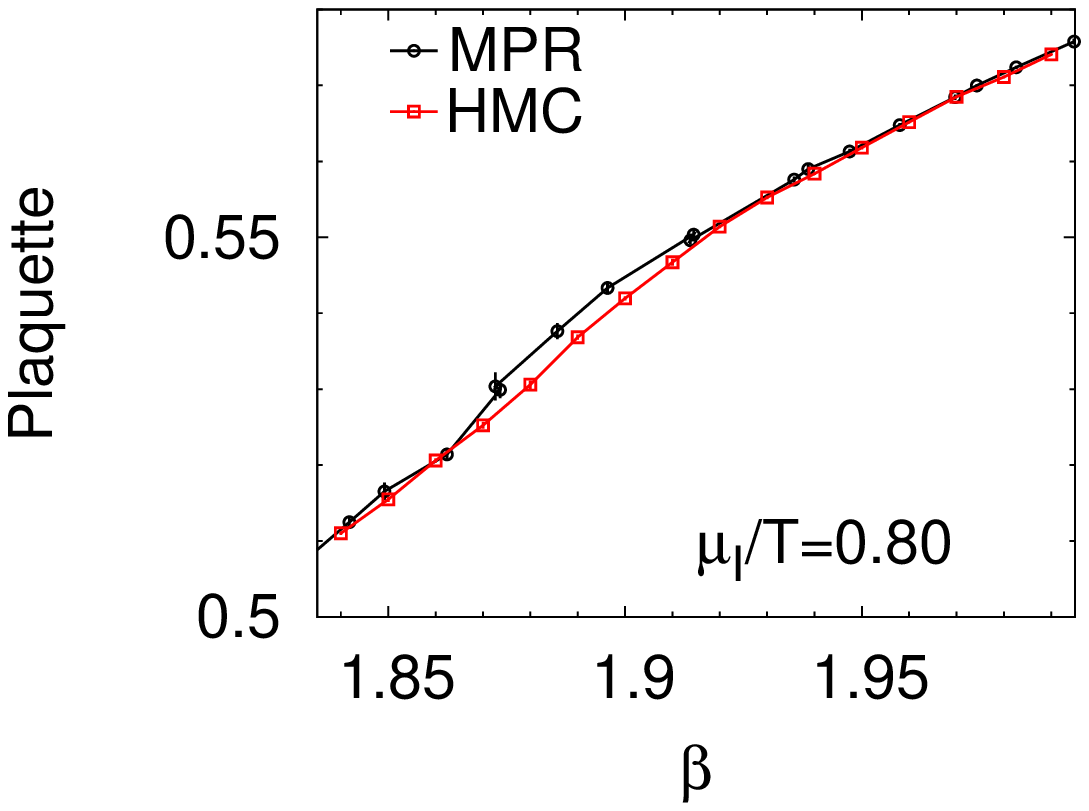}
\caption{Plaquette at imaginary chemical potential. 
The results for HMC were obtained from the direct simulations at $\mu_I$, 
while those for MPR were obtained from the simulations at $\mu=0$ with 
MPR method.}
\label{Sep252011Fig1}
\end{figure}
Since the comparison of MPR and Taylor expansion was done by using 
the same configurations, it may happen that both the methods are breakdown 
with same systematic errors. 
For further check, we consider the plaquette at imaginary chemical potential 
$\mu_I$ and compare the results with direct simulations. 
The consistency among several finite density lattice simulations 
was studied for staggered fermions in Ref.~\cite{Kratochvila:2005mk}.
The results are shown in Fig.~\ref{Sep252011Fig1}, where the data of direct 
simulations are taken from ~\cite{Nagata:2011yf}. 
MPR is almost consistent with the direct simulation up to  $\mu_I a=0.20$, although 
they are obtained from different configurations.
A small disagreement appears for $\mu_I a=0.20$ and it becomes larger for larger $\mu_I$. 
This agreement shows that the overlap problem caused by the real part of 
the reweighting factor is not severe up to $\mu_I/T=0.8$.
Note that the small error owes to the absence of the phase of the determinant at 
$\mu_I$ and that this consistency is irrelevant of the problem caused by the
imaginary part of $R$.

\subsection{Finite size effects}
\label{subsec3f}

\begin{figure}
\includegraphics[width=7cm]{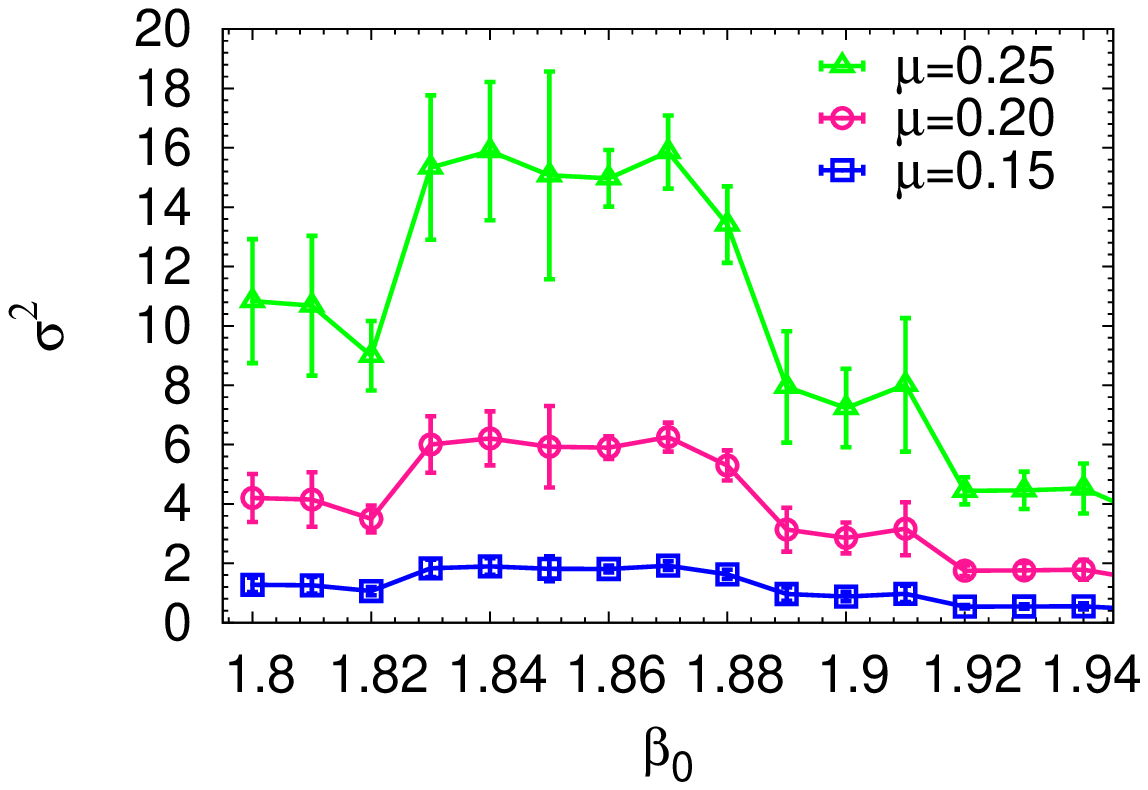}
\includegraphics[width=7cm]{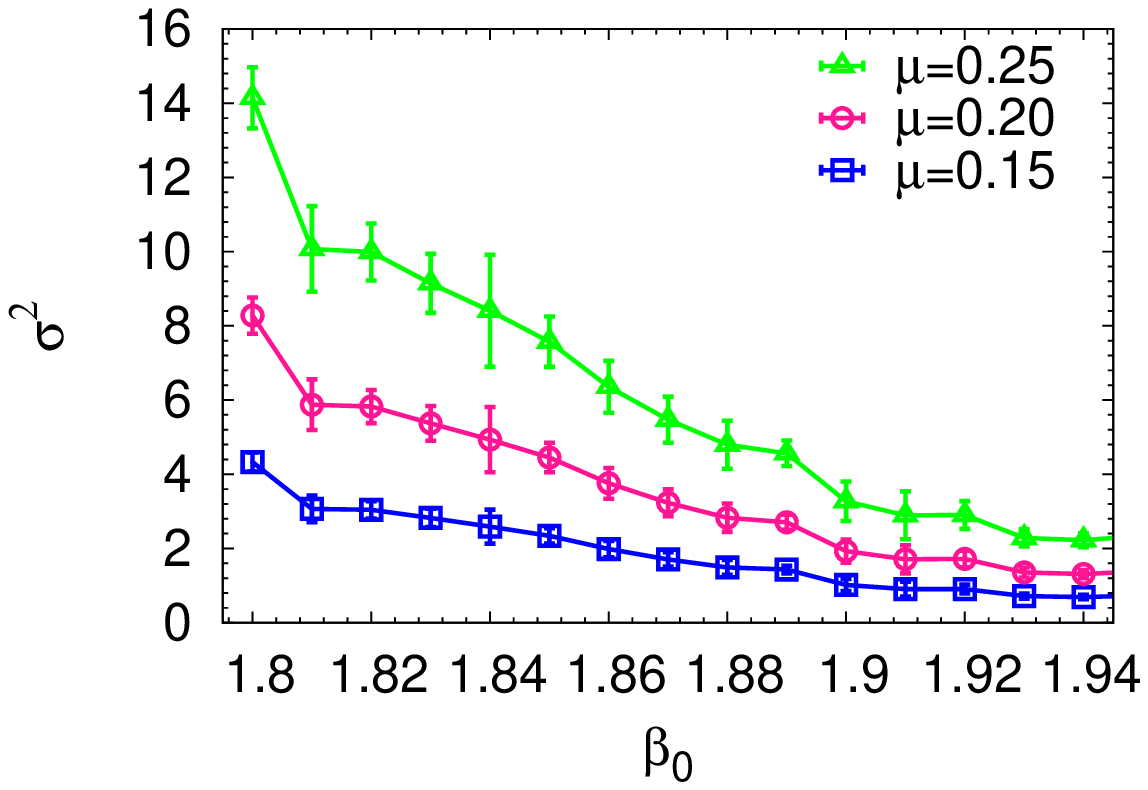}
\caption{The fluctuation of the quark determinant on $10^3\times 4$. 
$\sigma$ is given in the caption of Fig.~\ref{QDet2011Oct21Fig1}.
}
\label{Nov092011Fig1}
\end{figure}
Finally, we consider the finite size effects. 
The fluctuation of the quark determinant is shown for $N_s=8$ and 
$N_s=10$ in Figs.~\ref{QDet2011Oct21Fig1} and \ref{Nov092011Fig1}, 
where two calculations were performed in the same number of statistics.  
The fluctuations are almost proportional to the spatial volume $10^3/8^3 \sim 2$
for both the power and phase. 
This implies a well known result~\cite{deForcrand:2010ys} 
that the severity of the overlap problem is proportional to $O(\exp(V))$. 
In particular, the phase fluctuation goes over $\pi/2$ at about $\mu a \sim 0.15$
near and below $\betapc$, which imposes the applicable limit of MPR on this 
lattice size with the given statistics.

\begin{figure}
\includegraphics[width=7cm]{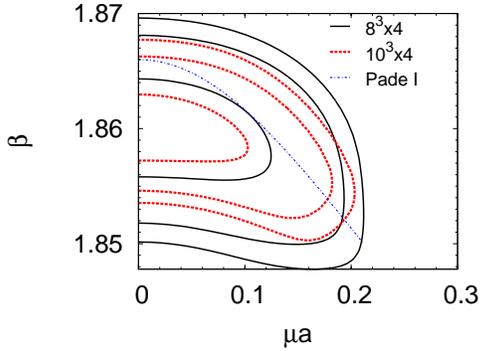}
\caption{
Contour lines of the fluctuation of the reweighting factor.
The solid and dashed lines are for $8^3\times 4$ and $10^3\times 4$. 
The pseudo critical line (Pad\'e I) is obtained in \cite{Nagata:2011yf}.
}
\label{Sep252011Fig3}
\end{figure}
In Fig.~\ref{Sep252011Fig3}, we show contour lines of $X$ for 
$8^3\times 4$ and $10^3\times 4$. 
The dotted line (Pad\'e I) shows the pseudo critical line obtained by 
the analytic continuation from imaginary chemical potential 
on the $8^3\times 4$~\cite{Nagata:2011yf}. 
The contour lines shrink due to the increase of $N_s$, 
the applicable range of the MPR becomes smaller for large lattice size. 
In order to extend the applicable range of MPR, it is required to 
increase statistics corresponding to the lattice size.

On the other hand, the shape of the contour line is similar for 
$N_s=8$ and $N_s=10$ in a sense that the fluctuation rapidly increases 
if the phase transition line is acrossed. 
It was shown in Ref.~\cite{Ejiri:2004yw} that in a system with a 
first order phase transition, the fluctuation of the reweighting factor
is minimum along the phase transition line, on a assumption 
that the fluctuation is dominated by the flip-flop between the two 
phases on the first order phase transition line. 
Although the phase transition is crossover, the fluctuation near $\Tpc$ 
is dominant by the one between hadron and QGP phases. 
Then, the direction of the reweighting line is insensitive to $N_s$. 
We have also confirmed that the EoS are not affected by 
the finite size effects up to  $\mu a = 0.2$, and number density and 
susceptibility up to $\mu a =0.10$. 
As long as we consider the parameter region with the small fluctuation, 
EoS, number density and susceptibility are insensitive to 
the lattice size, probably owing to the crossover nature of the deconfinement 
transition.

\section{Summary}

We have studied thermodynamical properties of QCD at nonzero  
quark chemical potential $\mu$ using the MPR and Taylor expansion methods with
a careful attention on the consistency of the MPR and Taylor expansion. 

Simulations were performed on the $8^3\times 4$ lattice 
with an intermediate quark mass region $m_{\rm ps}/m_{\rm V}\sim 0.8$
with the clover-improved Wilson fermion and RG-improved gauge action. 
The HMC simulation was done for 11 000 trajectories. 
Although the lattice size is small, the quark determinant was evaluated 
exactly by using the reduction formula for the Wilson fermion determinant. 
The eigenvalues of the reduced matrix were calculated for 400 configurations. 

Rapid fluctuation of the reweighting factor is known to cause the breakdown of MPR. 
To avoid the difficulty, we investigated the fluctuation of the reweighting factor. 
We have confirmed that the fluctuation of the reweighting factor is enough small up to 
$\mu a \sim 0.2$ both in the magnitude and phase. 
For the Taylor expansion, we evaluated the Taylor coefficients up to tenth order. 
Then, we have calculated the EoS, quark number density and quark number susceptibility. 
The MPR and Taylor expansion methods show a good agreement for the EoS and number 
density up to $\mu/T\sim 0.8$ and number susceptibility up to $\mu/T\sim 0.6$. 

One of the difficulty of the MPR method is the determination of the 
reweighting line, since it needs the calculation of the determinant 
for many parameter sets. 
We have derived the equation of the reweighting line and 
showed that the equation of the reweighting line is consistent with the 
fluctuation minimum condition for the calculation of the thermodynamical 
quantities. 
Using the equation of the reweighting line, one can avoid the determinant 
evaluation to search the fluctuation minimum line. 

To see how the obtained results are affected by finite size effects, 
we have compared $8^3\times 4$ and $10^3\times 4$. 
As expected, the fluctuation of the quark determinant increases as 
the volume becomes larger. In particular, the large fluctuation of the phase 
makes the applicable parameter range of MPR smaller. The phase fluctuation 
goes over $\pi/2$ for $\mu a \sim 0.15$ on the $10^3\times 4$ lattice. 
As long as we consider the parameter region with the small fluctuation, 
EoS, number density and susceptibility are insensitive to 
the lattice size, probably owing to the crossover nature of the deconfinement 
transition. 

The Taylor expansion and MPR methods have different advantage and difficulty.
The MPR method suffer from the fluctuation of the reweighting factor, while it is
free from truncation error of Taylor series. On the other hand, 
the Taylor expansion suffer from the truncation error, while it does not 
contain the reweighting factor.  
Thus, the obtained agreement between the two methods implies that the overlap problem 
for the MPR and truncation error for the Taylor expansion method are negligible for 
small $\mu$ and that the thermodynamical quantities are reliable for these errors. 

Although the present analysis is limited to small $\mu$ region, 
CEP may be located on a small or moderate $\mu$ region. 
The consistency observed here would be useful information for the studies of the CEP search.

\section*{Acknowledgment}
This work was supported by Grants-in-Aid for Scientific Research 20340055 and 20105003.
The simulation was performed on NEC SX-8R at RCNP, and NEC SX-9 at CMC, Osaka University.

\providecommand{\href}[2]{#2}\begingroup\raggedright\endgroup


\begin{thebibliography}{40}

\bibitem{deForcrand:2010ys}
P.~de~Forcrand, {\it {Simulating QCD at finite density}},  {\em PoS} {\bf
  LAT2009} (2009) 010, [\href{http://xxx.lanl.gov/abs/1005.0539}{{\tt
  arXiv:1005.0539}}].

\bibitem{Muroya:2003qs}
S.~Muroya, A.~Nakamura, C.~Nonaka, and T.~Takaishi, {\it {Lattice QCD at finite
  density: An Introductory review}},  {\em Prog.Theor.Phys.} {\bf 110} (2003)
  615--668, [\href{http://xxx.lanl.gov/abs/hep-lat/0306031}{{\tt
  hep-lat/0306031}}].

\bibitem{Schmidt:2006us}
C.~Schmidt, {\it {Lattice QCD at finite density}},  {\em PoS} {\bf LAT2006}
  (2006) 021, [\href{http://xxx.lanl.gov/abs/hep-lat/0610116}{{\tt
  hep-lat/0610116}}].

\bibitem{Ferrenberg:1988yz}
A.~Ferrenberg and R.~Swendsen, {\it {New Monte Carlo Technique for Studying
  Phase Transitions}},  {\em Phys.Rev.Lett.} {\bf 61} (1988) 2635--2638.

\bibitem{Barbour:1991vs}
I.~M. Barbour and A.~J. Bell, {\it {Complex zeros of the partition function for
  lattice QCD}},  {\em Nucl. Phys.} {\bf B372} (1992) 385--402.

\bibitem{Barbour:1997ej}
I.~M. Barbour, S.~E. Morrison, E.~G. Klepfish, J.~B. Kogut, and M.-P. Lombardo,
  {\it {Results on finite density QCD}},  {\em Nucl. Phys. Proc. Suppl.} {\bf
  60A} (1998) 220--234, [\href{http://xxx.lanl.gov/abs/hep-lat/9705042}{{\tt
  hep-lat/9705042}}].

\bibitem{Fodor:2001au}
Z.~Fodor and S.~Katz, {\it {A New method to study lattice QCD at finite
  temperature and chemical potential}},  {\em Phys.Lett.} {\bf B534} (2002)
  87--92, [\href{http://xxx.lanl.gov/abs/hep-lat/0104001}{{\tt
  hep-lat/0104001}}].

\bibitem{Fodor:2001pe}
Z.~Fodor and S.~Katz, {\it {Lattice determination of the critical point of QCD
  at finite T and mu}},  {\em JHEP} {\bf 0203} (2002) 014,
  [\href{http://xxx.lanl.gov/abs/hep-lat/0106002}{{\tt hep-lat/0106002}}].

\bibitem{Fodor:2002km}
Z.~Fodor, S.~Katz, and K.~Szabo, {\it {The QCD equation of state at nonzero
  densities: Lattice result}},  {\em Phys.Lett.} {\bf B568} (2003) 73--77,
  [\href{http://xxx.lanl.gov/abs/hep-lat/0208078}{{\tt hep-lat/0208078}}].

\bibitem{Fodor:2004nz}
Z.~Fodor and S.~Katz, {\it {Critical point of QCD at finite T and mu, lattice
  results for physical quark masses}},  {\em JHEP} {\bf 0404} (2004) 050,
  [\href{http://xxx.lanl.gov/abs/hep-lat/0402006}{{\tt hep-lat/0402006}}].

\bibitem{Fodor:2009ax}
Z.~Fodor and S.~Katz, {\it {The Phase diagram of quantum chromodynamics}},
  \href{http://xxx.lanl.gov/abs/0908.3341}{{\tt arXiv:0908.3341}}.

\bibitem{Ejiri:2004yw}
S.~Ejiri, {\it {Remarks on the multiparameter reweighting method for the study
  of lattice QCD at nonzero temperature and density}},  {\em Phys.Rev.} {\bf
  D69} (2004) 094506, [\href{http://xxx.lanl.gov/abs/hep-lat/0401012}{{\tt
  hep-lat/0401012}}].

\bibitem{Allton:2002zi}
C.~Allton, S.~Ejiri, S.~Hands, O.~Kaczmarek, F.~Karsch, {\em et.~al.}, {\it
  {The QCD thermal phase transition in the presence of a small chemical
  potential}},  {\em Phys.Rev.} {\bf D66} (2002) 074507,
  [\href{http://xxx.lanl.gov/abs/hep-lat/0204010}{{\tt hep-lat/0204010}}].

\bibitem{Allton:2003vx}
C.~Allton, S.~Ejiri, S.~Hands, O.~Kaczmarek, F.~Karsch, {\em et.~al.}, {\it
  {The Equation of state for two flavor QCD at nonzero chemical potential}},
  {\em Phys.Rev.} {\bf D68} (2003) 014507,
  [\href{http://xxx.lanl.gov/abs/hep-lat/0305007}{{\tt hep-lat/0305007}}].

\bibitem{Allton:2005gk}
C.~Allton, M.~Doring, S.~Ejiri, S.~Hands, O.~Kaczmarek, {\em et.~al.}, {\it
  {Thermodynamics of two flavor QCD to sixth order in quark chemical
  potential}},  {\em Phys.Rev.} {\bf D71} (2005) 054508,
  [\href{http://xxx.lanl.gov/abs/hep-lat/0501030}{{\tt hep-lat/0501030}}].

\bibitem{Gavai:2004sd}
R.~Gavai and S.~Gupta, {\it {The Critical end point of QCD}},  {\em Phys.Rev.}
  {\bf D71} (2005) 114014, [\href{http://xxx.lanl.gov/abs/hep-lat/0412035}{{\tt
  hep-lat/0412035}}].

\bibitem{Ejiri:2009hq}
{\bf WHOT-QCD Collaboration} Collaboration, S.~Ejiri {\em et.~al.}, {\it
  {Equation of State and Heavy-Quark Free Energy at Finite Temperature and
  Density in Two Flavor Lattice QCD with Wilson Quark Action}},  {\em
  Phys.Rev.} {\bf D82} (2010) 014508,
  [\href{http://xxx.lanl.gov/abs/0909.2121}{{\tt arXiv:0909.2121}}].

\bibitem{Csikor:2004ik}
F.~Csikor, G.~Egri, Z.~Fodor, S.~Katz, K.~Szabo, {\em et.~al.}, {\it {Equation
  of state at finite temperature and chemical potential, lattice QCD results}},
   {\em JHEP} {\bf 0405} (2004) 046,
  [\href{http://xxx.lanl.gov/abs/hep-lat/0401016}{{\tt hep-lat/0401016}}].

\bibitem{Borici:2004bq}
A.~Borici, {\it {Reweighting with stochastic determinants}},  {\em Prog. Theor.
  Phys. Suppl.} {\bf 153} (2004) 335--339.

\bibitem{Nagata:2010xi}
K.~Nagata and A.~Nakamura, {\it {Wilson Fermion Determinant in Lattice QCD}},
  {\em Phys.Rev.} {\bf D82} (2010) 094027,
  [\href{http://xxx.lanl.gov/abs/1009.2149}{{\tt arXiv:1009.2149}}].

\bibitem{Alexandru:2010yb}
A.~Alexandru and U.~Wenger, {\it {QCD at non-zero density and canonical
  partition functions with Wilson fermions}},  {\em Phys.Rev.} {\bf D83} (2011)
  034502, [\href{http://xxx.lanl.gov/abs/1009.2197}{{\tt arXiv:1009.2197}}].

\bibitem{Ferrenberg:1989ui}
A.~M. Ferrenberg and R.~H. Swendsen, {\it {Optimized Monte Carlo analysis}},
  {\em Phys.Rev.Lett.} {\bf 63} (1989) 1195--1198.

\bibitem{Kratochvila:2005mk}
S.~Kratochvila and P.~de~Forcrand, {\it {The canonical approach to finite
  density QCD}},  {\em PoS} {\bf LAT2005} (2006) 167,
  [\href{http://xxx.lanl.gov/abs/hep-lat/0509143}{{\tt hep-lat/0509143}}].

\bibitem{deForcrand:2006ec}
P.~de~Forcrand and S.~Kratochvila, {\it {Finite density QCD with a canonical
  approach}},  {\em Nucl. Phys. Proc. Suppl.} {\bf 153} (2006) 62--67,
  [\href{http://xxx.lanl.gov/abs/hep-lat/0602024}{{\tt hep-lat/0602024}}].

\bibitem{Nakagawa:2011eu}
{\bf WHOT-QCD collaboration} Collaboration, Y.~Nakagawa {\em et.~al.}, {\it
  {Histogram method in finite density QCD with phase quenched simulations}},
  \href{http://xxx.lanl.gov/abs/1111.2116}{{\tt arXiv:1111.2116}}.

\bibitem{Gibbs:1986hi}
P.~E. Gibbs, {\it {THE FERMION PROPAGATOR MATRIX IN LATTICE QCD}},  {\em Phys.
  Lett.} {\bf B172} (1986) 53.

\bibitem{Hasenfratz:1991ax}
A.~Hasenfratz and D.~Toussaint, {\it {Canonical ensembles and nonzero density
  quantum chromodynamics}},  {\em Nucl. Phys.} {\bf B371} (1992) 539--549.

\bibitem{Schaefer:2009st}
B.-J. Schaefer, M.~Wagner, and J.~Wambach, {\it {QCD thermodynamics with
  effective models}},  {\em PoS} {\bf CPOD2009} (2009) 017,
  [\href{http://xxx.lanl.gov/abs/0909.0289}{{\tt arXiv:0909.0289}}].

\bibitem{Nagata:2011yf}
K.~Nagata and A.~Nakamura, {\it {Imaginary Chemical Potential Approach for the
  Pseudo-Critical Line in the QCD Phase Diagram with Clover-Improved Wilson
  Fermions}},  {\em Phys.Rev.} {\bf D83} (2011) 114507,
  [\href{http://xxx.lanl.gov/abs/1104.2142}{{\tt arXiv:1104.2142}}].

\end{thebibliography}

\end{document}